\documentclass{article}

\def\xxinput#1{\input#1}

\xxinput{vsolj02.sty}

\usepackage[dvipdfmx]{graphicx}
\usepackage[comma,colon]{natbib}

\usepackage[OT2,T1]{fontenc}

\def\cite{\citealt}
\setcitestyle{aysep={}}

\def\commenta{$^*$}
\def\commentb{$^\dagger$}

\newcounter{author}
\def\altaffilmark#1{$^{#1}$}
\def\altaffiltext#1{$^{#1}$\,}

\def\authorcount#1#2{{\refstepcounter{author}\label{#1}
                     \altaffiltext{\ref{#1}}{#2}}}

\begin{document}

\begin{center}

\title{CM Mic and other ER UMa stars showing standstills}

\author{
        Taichi~Kato\altaffilmark{\ref{affil:Kyoto}},
        Naoto~Kojiguchi\altaffilmark{\ref{affil:Kyoto}}}

\authorcount{affil:Kyoto}{
     Department of Astronomy, Kyoto University, Sakyo-ku,
     Kyoto 606-8502, Japan}
\email{tkato@kusastro.kyoto-u.ac.jp, kojiguchi@kusastro.kyoto-u.ac.jp}

\end{center}

\begin{abstract}
\xxinput{abst.inc}
\end{abstract}

\section{Introduction}

   ER UMa stars are a small group of SU UMa-type dwarf novae
with very short (typically shorter than $\sim$80~d)\footnote{
   Short supercycles are a result of high mass-transfer rates.
   As \citet{osa95eruma} showed, however, supercycles can become
   longer when superoutbursts are long and their durations comprise
   most of a supercycle, as we will see in RZ LMi and
   CM Mic in 2020.
   This situation should be distinguished from systems
   with long supercycles but with short superoutbursts
   (i.e. ordinary SU UMa stars).
}
supercycles and frequent normal outbursts
\citep{kat95eruma,rob95eruma,pat95v1159ori,kat99erumareview}
[normal outbursts, however, can be suppressed, see
e.g., \citet{zem13eruma,ohs14eruma} and normal outbursts sometimes
may not occur frequently].
A small number of ER UMa stars and related systems
have been recorded to show standstills,
which are the defining characteristic of the Z Cam-type
dwarf nova [for general information of cataclysmic variables (CVs)
and dwarf novae, see e.g., \citet{war95book}].

   Such objects already reported in the literature are
as follows.
\begin{itemize}
\item
RZ LMi showed long-lasting superoutbursts
in 2016--2017 \citep{kat16rzlmi}.  The general light curve
and the durations of the flat parts are similar to
the recently reported standstill of the AM CVn star CR Boo
\citep{kat23crboo}.  It appears that the phenomena recorded in
RZ LMi can be interpreted as standstills.
\item
BK Lyn was originally considered as a novalike object
(a CV with a thermally stable disk) \citep{ski93bklyn,rin96bklyn}.
The object was found to be in an ER UMa state in 2011--2012
\citep{kem12bklynsass,pat13bklyn,Pdot4}.
The object was probably in this state already in 2005,
but not as early as in 2002 \citep{Pdot4}.
It returned to the novalike state in 2013 \citep{Pdot4}
and has been in this state since then.
This object is probably essentially a novalike object
with a transient ER UMa state.
\item
NY Ser is an SU UMa-type dwarf nova in the period gap
\citep{nog98nyser,skl18nyser}.  Although supercycles
were not as regular as in typical ER UMa stars, the shortest
interval of confirmed superoutburst was 97~d (1996 April
and July--Augurst, see \cite{nog98nyser}).  This object
also shows long normal outbursts \citep{pav14nyser} and
the previous report of a 85~d supercycle [\citet{nog98nyser},
adopted in \citet{RKCat}] is probably an underestimate.
Nevertheless, we include this object here as an object related
to ER UMa stars.  This object showed two standstills and
superoutbursts arising from them in 2018 \citep{kat19nyser}.
This was the first demonstration that the disk radius
can increase during standstills.
\item
MGAB-V859 and ZTF18abgjsdg were reported in
\citet{kat21mgabv859ztf18abgjsdg} showing standstill phases
in addition to ER UMa-type behavior.
\item
WFI J161953.3$+$031909 is an eclipsing object showing both
a standstill and ER UMa states \citep{kat22j1619}.
\end{itemize}

   ER UMa stars are considered to have higher mass-transfer rates
than expected from the standard evolutionary sequence
of CVs, and are suspected to be descendants of old novae
(see e.g., \cite{pat13bklyn}).  The presence of standstills
also favors a high mass-transfer rate as required by
the disk instability theory (see e.g., \cite{osa96review}).

   ER UMa stars showing standstills have additional implications:
they are probably a result of variable mass-transfer rates
and how such standstills evolve provides a clue in understanding
the evolution of the disk radius or the angular momentum
in the accretion disk \citep{kat19nyser}, which is still
an unsolved problem (e.g., \cite{kim20iwandmodel}).

   In this paper, we show that CM Mic (=EC 20335$-$4332)
is yet another member of this small group of CVs.
This object was initially selected as a DC white dwarf
(CS 22943-206: \cite{bee92hotstars}).  \citet{che01ECCV}
classified it as a possible dwarf nova with a range
of $B$=14.9--15.9.  \citet{NameList77} gave a variable star
name CM Mic and classified it as a novalike.
This object did not receive much attention and it was only
2018 when the object was classified as a Z Cam star based
on the All-Sky Automated Survey for Supernovae (ASAS-SN,
\cite{ASASSN}, \cite{koc17ASASSNLC}) observations
(T. Kato, vsnet-chat 8095\footnote{
  $<$http://ooruri.kusastro.kyoto-u.ac.jp/mailarchive/vsnet-chat/8095$>$.
}).  Using ASAS-SN and Transiting Exoplanet Survey Satellite (TESS)
observations \citep{ric15TESS}\footnote{
  $<$https://tess.mit.edu/observations/$>$.
  The full light-curve
  is available at the Mikulski Archive for Space Telescope
  (MAST, $<$http://archive.stsci.edu/$>$).
}, this object was also identified as an ER UMa star
(N. Kojiguchi, vsnet-chat 8992\footnote{
  $<$http://ooruri.kusastro.kyoto-u.ac.jp/mailarchive/vsnet-chat/8992$>$.
}).

\section{Data analysis}

   We used ASAS-SN observations and Asteroid Terrestrial-impact
Last Alert System (ATLAS: \cite{ATLAS}) forced photometry
\citep{shi21ALTASforced} to examine the long-term behavior.

   We used TESS observations, which recorded the initial part of
the 2020 July superoutburst, to analyze superhumps.
Superhumps maxima were determined using the template fitting
method introduced in \citet{Pdot} after removing outburst trends
by locally-weighted polynomial regression (LOWESS: \cite{LOWESS}).
The periods were determined using the phase dispersion minimization
(PDM: \cite{PDM}) method, whose errors were
estimated by the methods of \citet{fer89error,Pdot2}.
We also used Zwicky Transient Facility (ZTF: \cite{ZTF})\footnote{
   The ZTF data can be obtained from IRSA
$<$https://irsa.ipac.caltech.edu/Missions/ztf.html$>$
using the interface
$<$https://irsa.ipac.caltech.edu/docs/program\_interface/ztf\_api.html$>$
or using a wrapper of the above IRSA API
$<$https://github.com/MickaelRigault/ztfquery$>$.
} data for analysis of other objects given in section
\ref{sec:other}.

\section{Results}

\subsection{Long-term behavior}\label{sec:longterm}

\begin{figure*}
\begin{center}
\includegraphics[width=16cm]{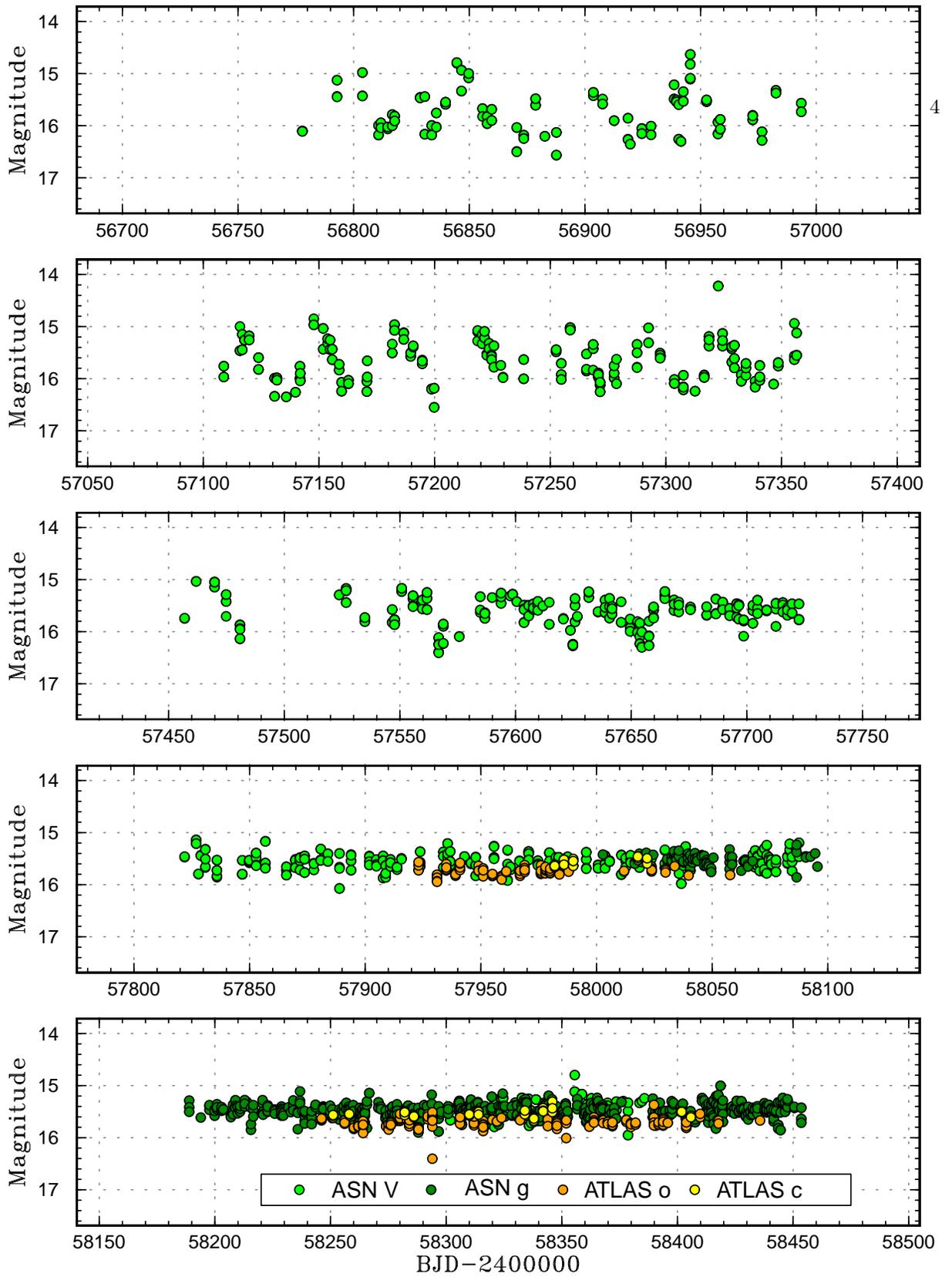}
\caption{
   Light curve of CM Mic in 2014--2018.  The object showed
outbursts with a cycle length of $\sim$35~d in 2015
(second panel).  This pattern became weaker in 2016 (third panel).
The object stayed in standstill in 2017 and 2018
(fourth and fifth panels).
ASN refer to ASAS-SN observations.
}
\label{fig:lc1}
\end{center}
\end{figure*}

\begin{figure*}
\begin{center}
\includegraphics[width=16cm]{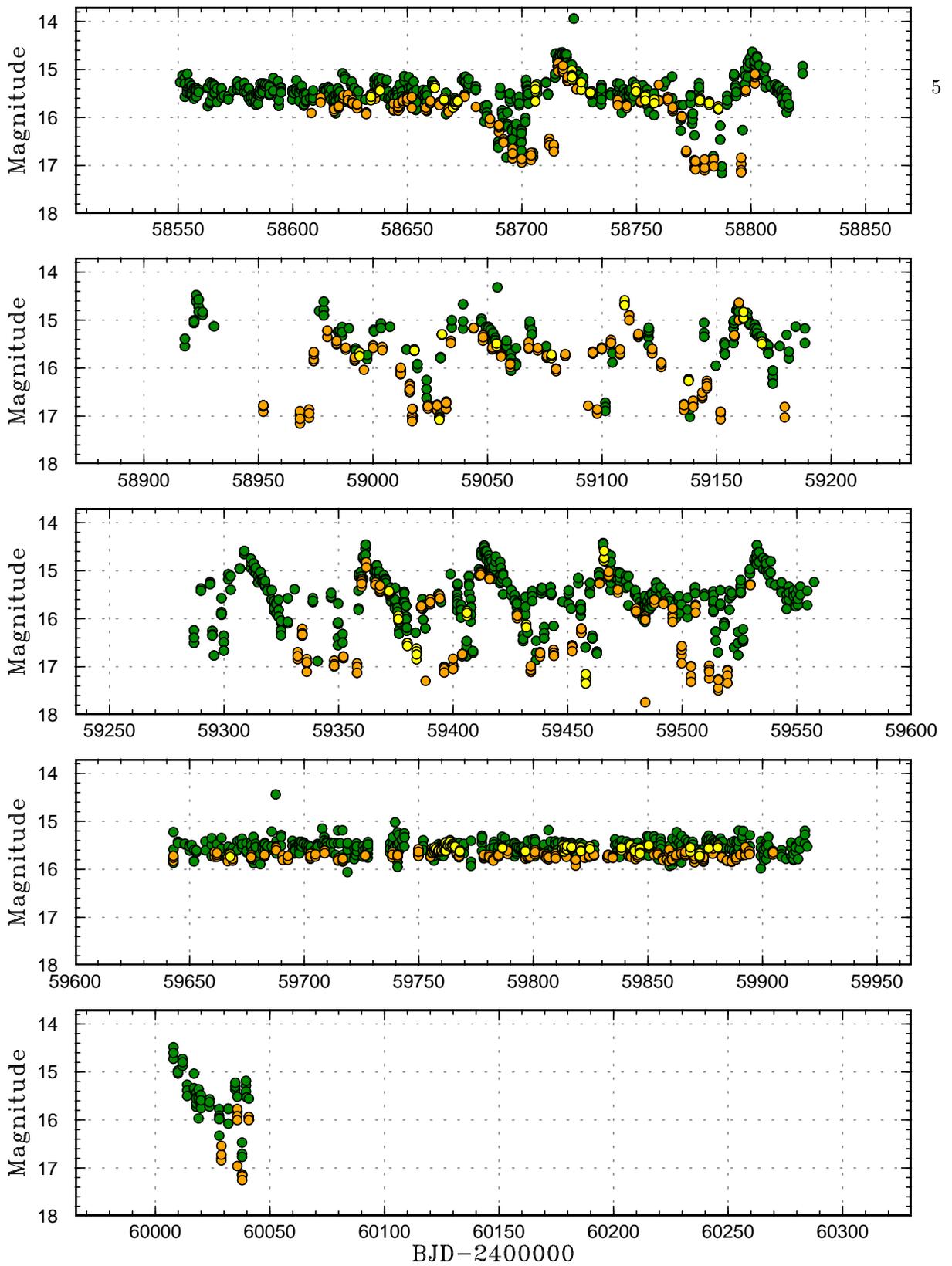}
\caption{
   Light curve of CM Mic in 2019--2023.
The symbols are the same as in figure \ref{fig:lc1}.
}
\label{fig:lc2}
\end{center}
\end{figure*}

   Long-term light curves based on ASAS-SN and ATLAS data are
shown in figures \ref{fig:lc1} and \ref{fig:lc2}.

   In figure \ref{fig:lc1}, the object showed
outbursts with a cycle length of $\sim$35~d in 2015 
(second panel).  This pattern became weaker in 2016 (third panel).
The object stayed in standstill in 2017 and 2018
(fourth and fifth panels).
The behavior in 2014 (first panel) was likely similar to 2015,
but was uncertain due to the limited quality of the ASAS-SN data.

\begin{figure*}
\begin{center}
\includegraphics[width=16cm]{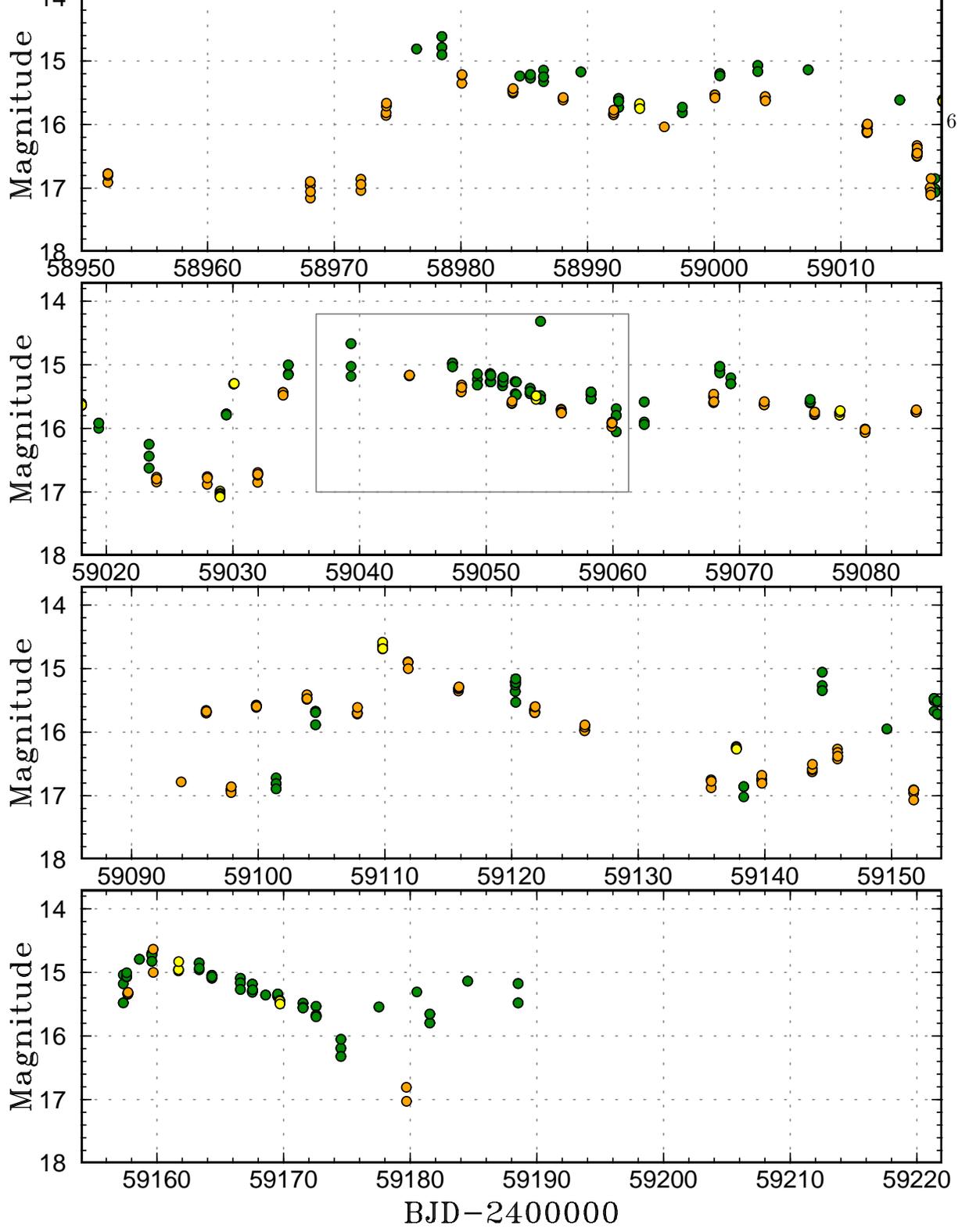}
\caption{
   ER UMa state of CM Mic in 2020.
The symbols are the same as in figure \ref{fig:lc1}.
The first and second superoutbursts (first and second panels,
respectively) were long and were associated with brightening
in the later phase.
The grey box indicates the range of figure \ref{fig:humpall}.
}
\label{fig:lc3}
\end{center}
\end{figure*}

\begin{figure*}
\begin{center}
\includegraphics[width=16cm]{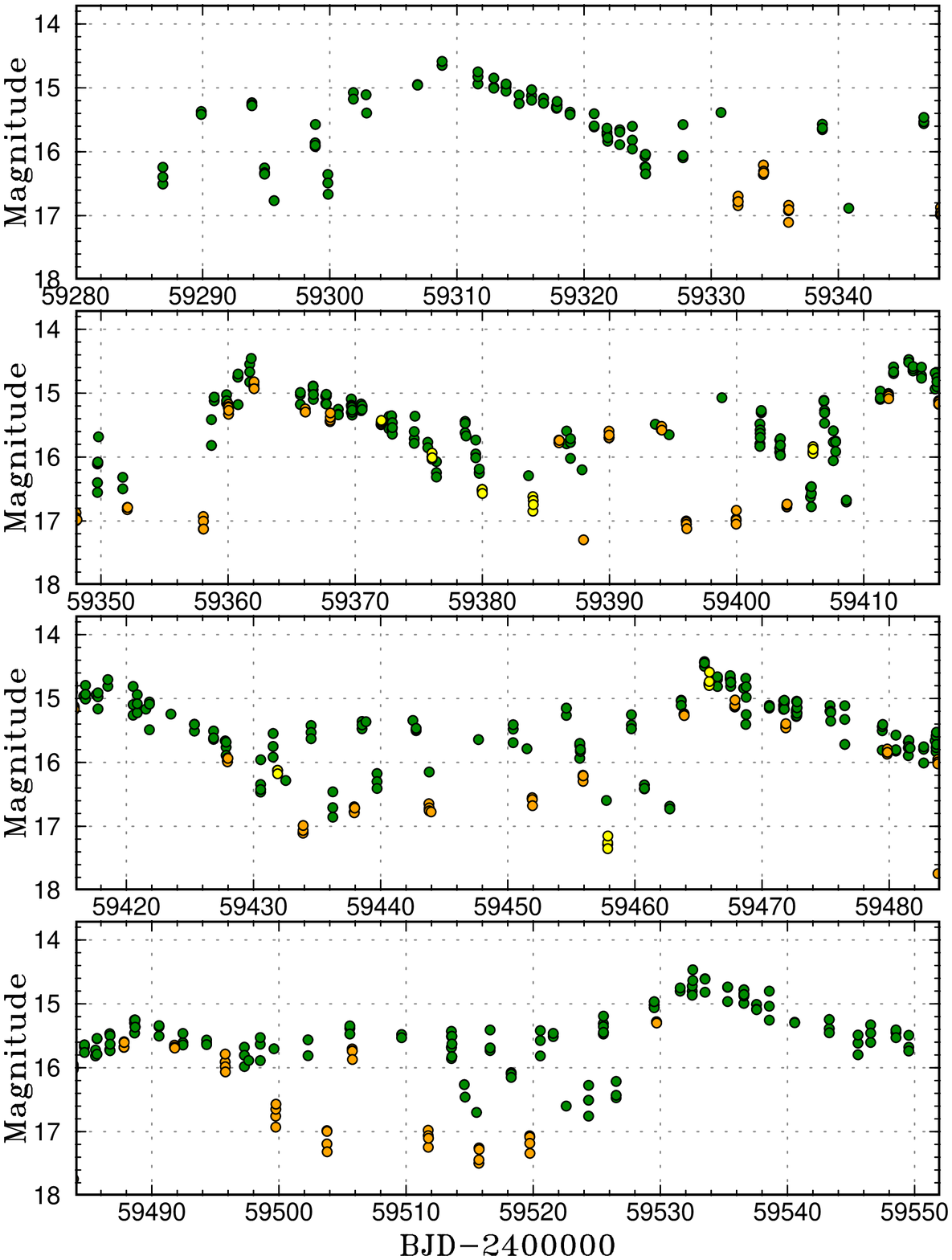}
\caption{
   ER UMa state of CM Mic in 2021.
The symbols are the same as in figure \ref{fig:lc1}.
Supercycles became shorter and more regular than in 2020.
}
\label{fig:lc4}
\end{center}
\end{figure*}

   In figure \ref{fig:lc2}, the object was initially in standstill
(first panel) with some hint of low-amplitude oscillation with
a period of 20--30~d.  The object entered a dwarf nova state
after BJD 2458680 (2019 July 15), followed by a superoutburst
on BJD 2458714 (2019 August 18)\footnote{
   We assume that all long outbursts in this object are superoutbursts
   based on the detection of superhumps during one of
   long outbursts in the TESS data (subsection \ref{sec:tess}).
}.  This superoutburst lasted
long ($\sim$55~d), as in long superoutbursts or superoutbursts
with a standstill in RZ LMi \citep{kat16rzlmi}.
The supercycle was 83~d.  In 2020 (second panel), the object
showed shorter supercycles (49--70~d) with some structures
in the fading part of the superhumps (such as at BJD 2459000 and
2459068).  In 2021 (third panel), the object showed a more
typical ER UMa-type variation with many normal outbursts
but with variable supercycles (53--67~d) and variable shapes
of superoutbursts.
The superoutburst starting on BJD 2459465 (2021 September 7)
showed brightening in the later part (BJD 2459488,
2021 October 1) as in 2020.
The object was in standstill again in 2022 (fourth panel).
It has been in an ER UMa state in 2023 again (fifth panel).
The ER UMa-type behavior in 2020 and 2021 is shown in more detail
in an expanded scale in figures \ref{fig:lc3} and \ref{fig:lc4}.

\subsection{TESS observations}\label{sec:tess}

\begin{figure*}
\begin{center}
\includegraphics[width=16cm]{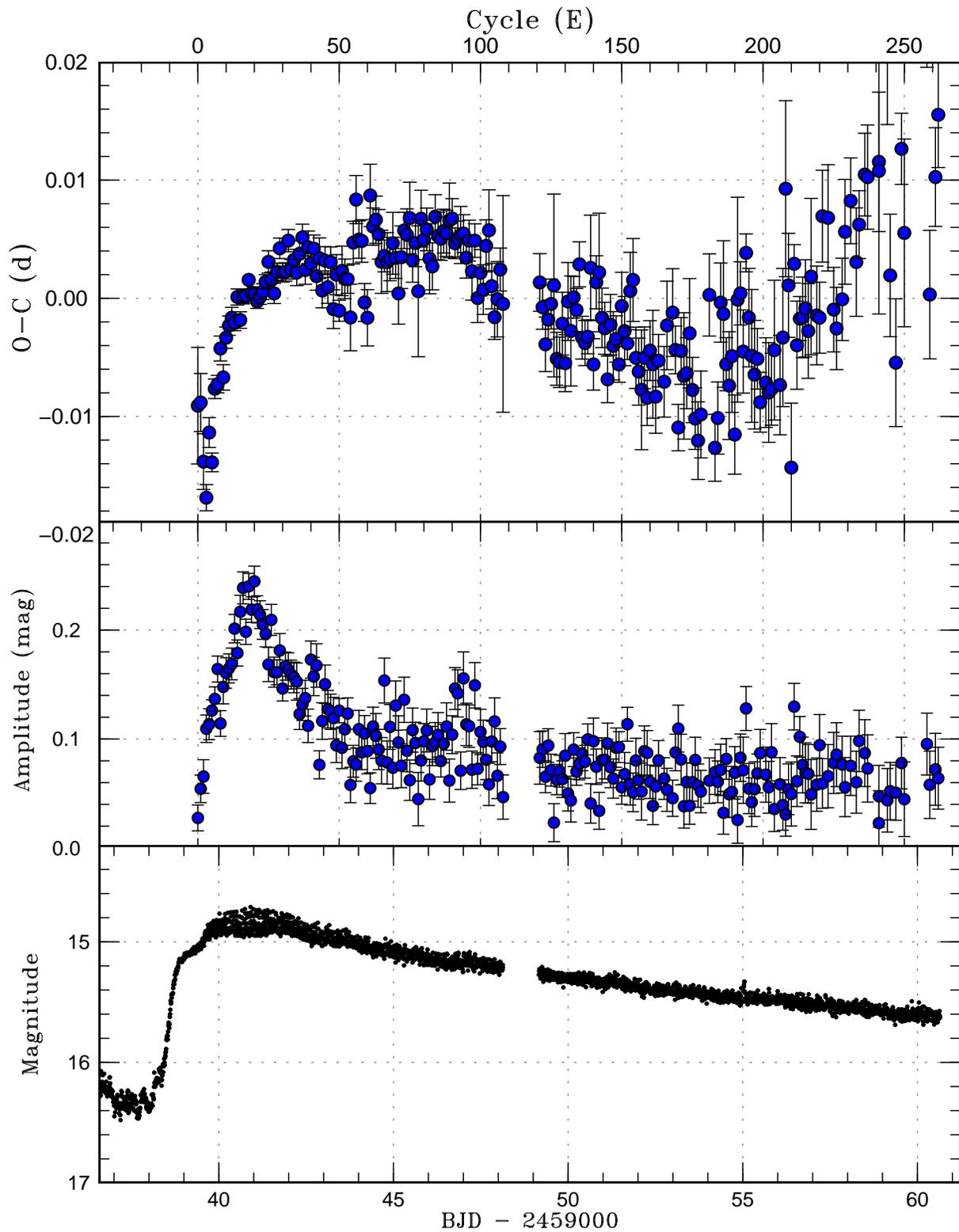}
\caption{
  TESS observations of superhumps of CM Mic during the 2020 July
  superoutburst.
  (Upper:) $O-C$ variation.  The ephemeris of
  BJD(max) = 2459039.4000$+$0.08090$E$ was used.
  The data are in table \ref{tab:shmax}.
  (Middle:) Superhump amplitude.
  (Lower:) TESS light curve.
  The data were binned to 0.008~d.
}
\label{fig:humpall}
\end{center}
\end{figure*}

\begin{figure*}
\begin{center}
\includegraphics[width=14cm]{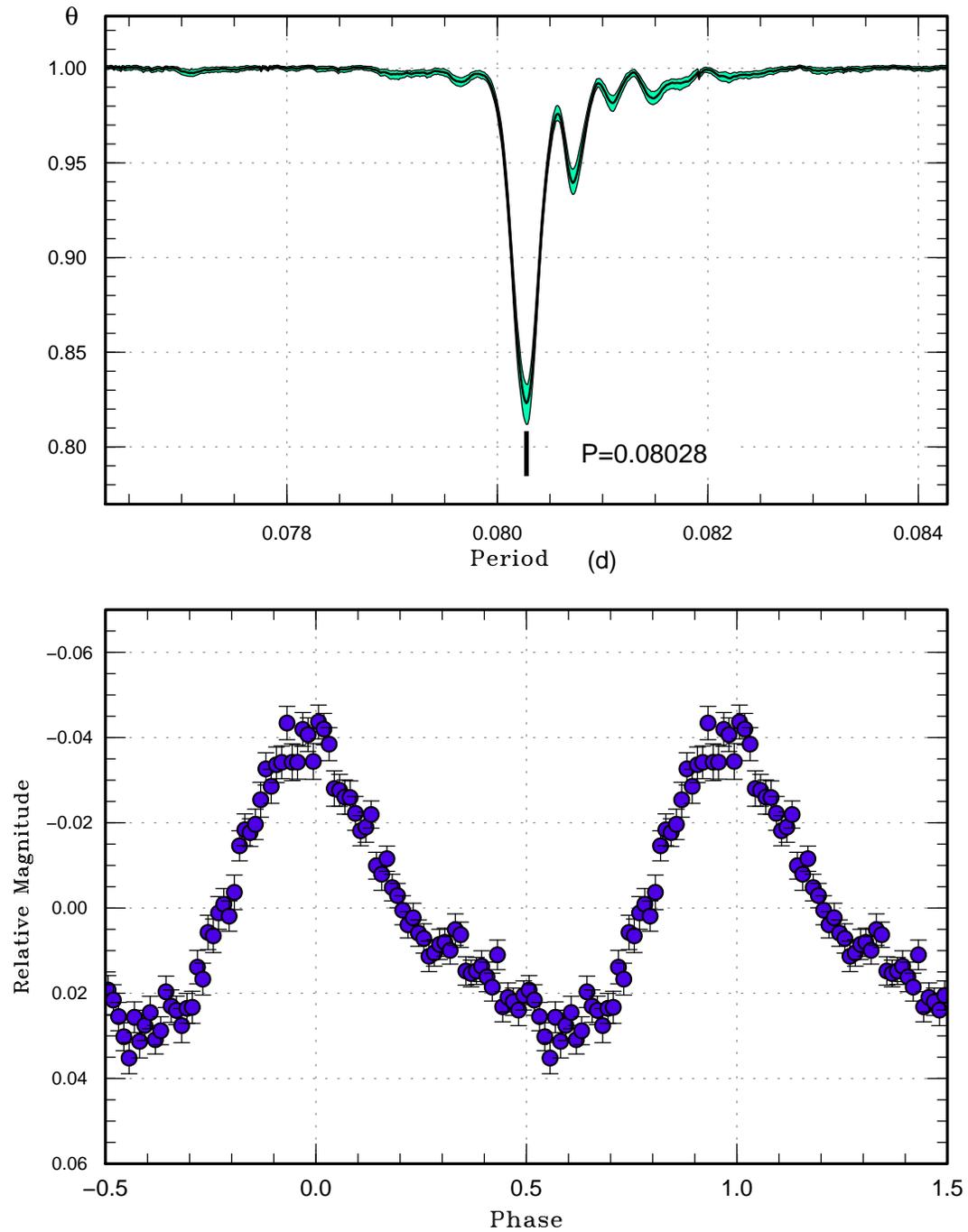}
\caption{
   Mean superhump profile of CM Mic from TESS data
   (range after BJD 2459039.4).
   (Upper): PDM analysis.  The bootstrap result using
   randomly contain 50\% of observations is shown as
   a form of 90\% confidence intervals in the resultant 
   $\theta$ statistics.
   (Lower): Phase plot.
}
\label{fig:pdm}
\end{center}
\end{figure*}

   The times of superhump maxima are listed in table
\ref{tab:shmax}.  The $O-C$ variation together with
the variation of superhump amplitudes and the light curve
is shown in figure \ref{fig:humpall}.  The peak of
the superhump amplitudes occurred around $E$=19.
TESS observations covered superhump stages A and B
[see \citet{Pdot,kat22v844her} for superhump stages].
Using the linear part 2$\le E \le$12,
we obtained a period of 0.0817(2)~d for stage A superhumps.
It was somewhat ambiguous to define when stage B started.
Using the range 30$\le E \le$262, the mean superhump period
($P_{\rm SH}$) was 0.080251(6)~d and the period
derivative ($P_{\rm dot} = \dot{P}/P$) was $+$2.0(2) $\times$ 10$^{-5}$.
Using the range 50$\le E \le$262, these values were
$P_{\rm SH}$ = 0.080253(8)~d and
$P_{\rm dot}$ = $+$2.9(3) $\times$ 10$^{-5}$.
The small positive $P_{\rm dot}$ was similar to
$+$4(2) $\times$ 10$^{-5}$ for ER UMa \citep{Pdot} and
$+$3(3) $\times$ 10$^{-5}$ for BK Lyn \citep{Pdot4}.
As shown in \citet{kat13qfromstageA,kat22stageA},
the mass ratio can be determined from the stage A
superhump period if the orbital period is known.
In the case of CM Mic, the orbital signal was not detected
in the available TESS data and this is a future task possibly
requiring a radial-velocity study.
A PDM analysis and the mean profile of superhumps from
TESS observations are shown in figure \ref{fig:pdm}.
The mean period (entire data with stages A and B combined)
was 0.080276(5)~d.

\begin{figure*}
\begin{center}
\includegraphics[width=16cm]{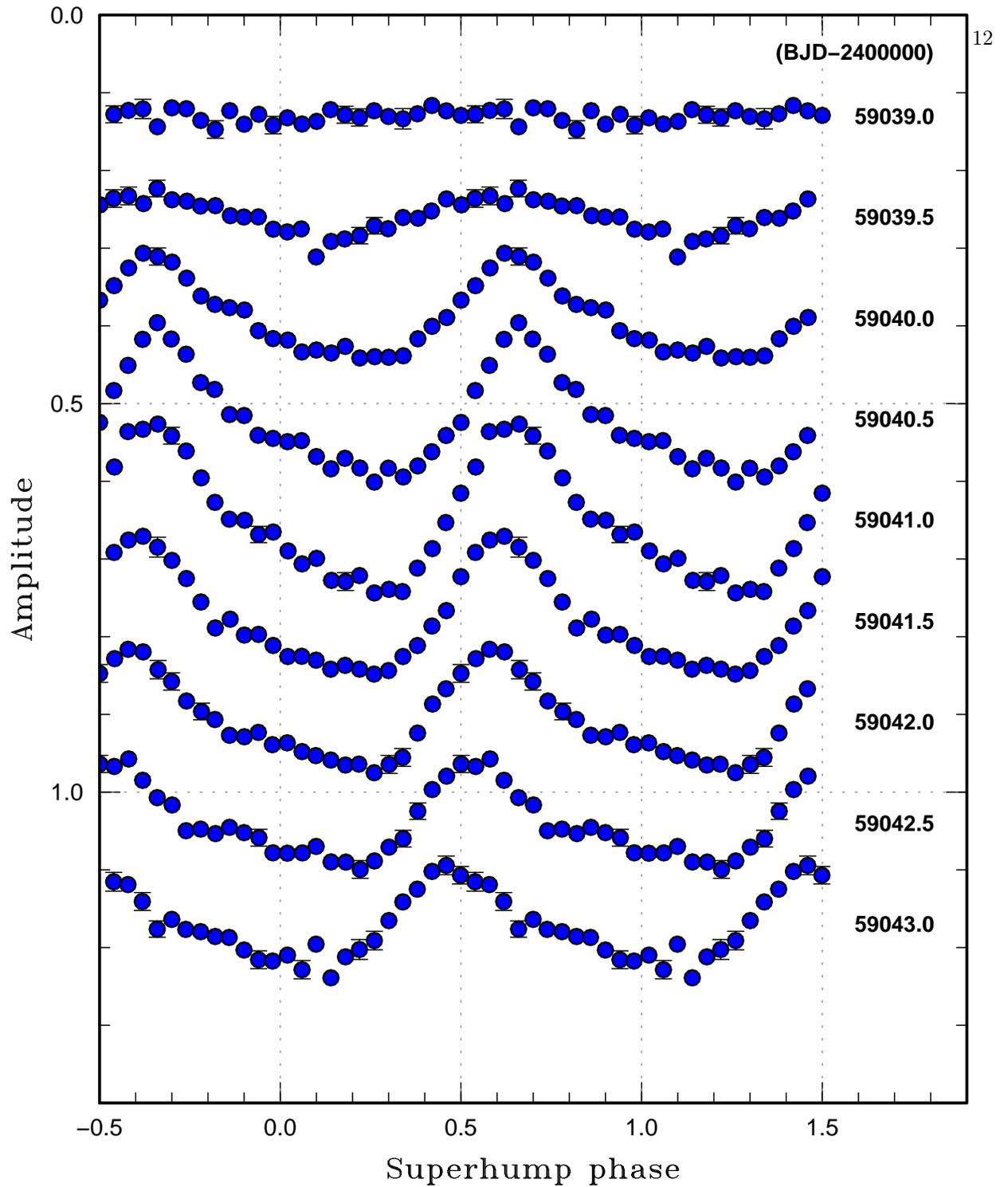}
\caption{
  Variation of profiles of superhumps in the growing phase (stage A)
  to the initial part of stage B.
  The zero phase and superhump period were defined as BJD 2459039.4000
  and 0.08090~d (same as in figure \ref{fig:humpall}).
  0.5-d segments were used whose centers are shown on the right side
  of the figure.
  The superhump period was long up to BJD 2459040.5 and then
  became short.
}
\label{fig:prof}
\end{center}
\end{figure*}

\begin{figure*}
\begin{center}
\includegraphics[width=16cm]{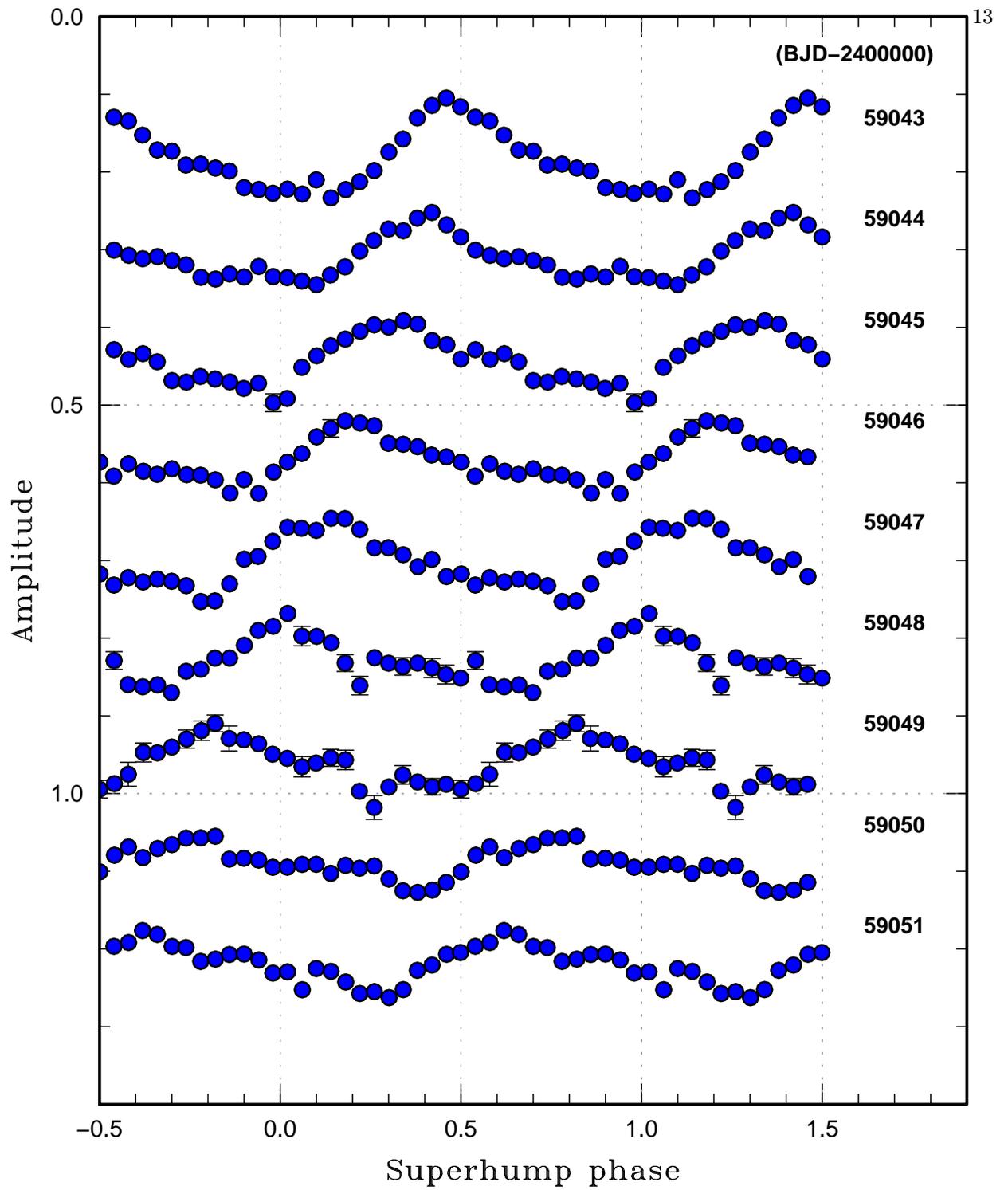}
\caption{
  Variation of profiles of superhumps in the early part of
  stage B.  1-d segments were used whose centers are shown on the
  right side of the figure.
  The zero phase and superhump period were
  the same as in figure \ref{fig:prof}.
}
\label{fig:prof2}
\end{center}
\end{figure*}

\begin{figure*}
\begin{center}
\includegraphics[width=16cm]{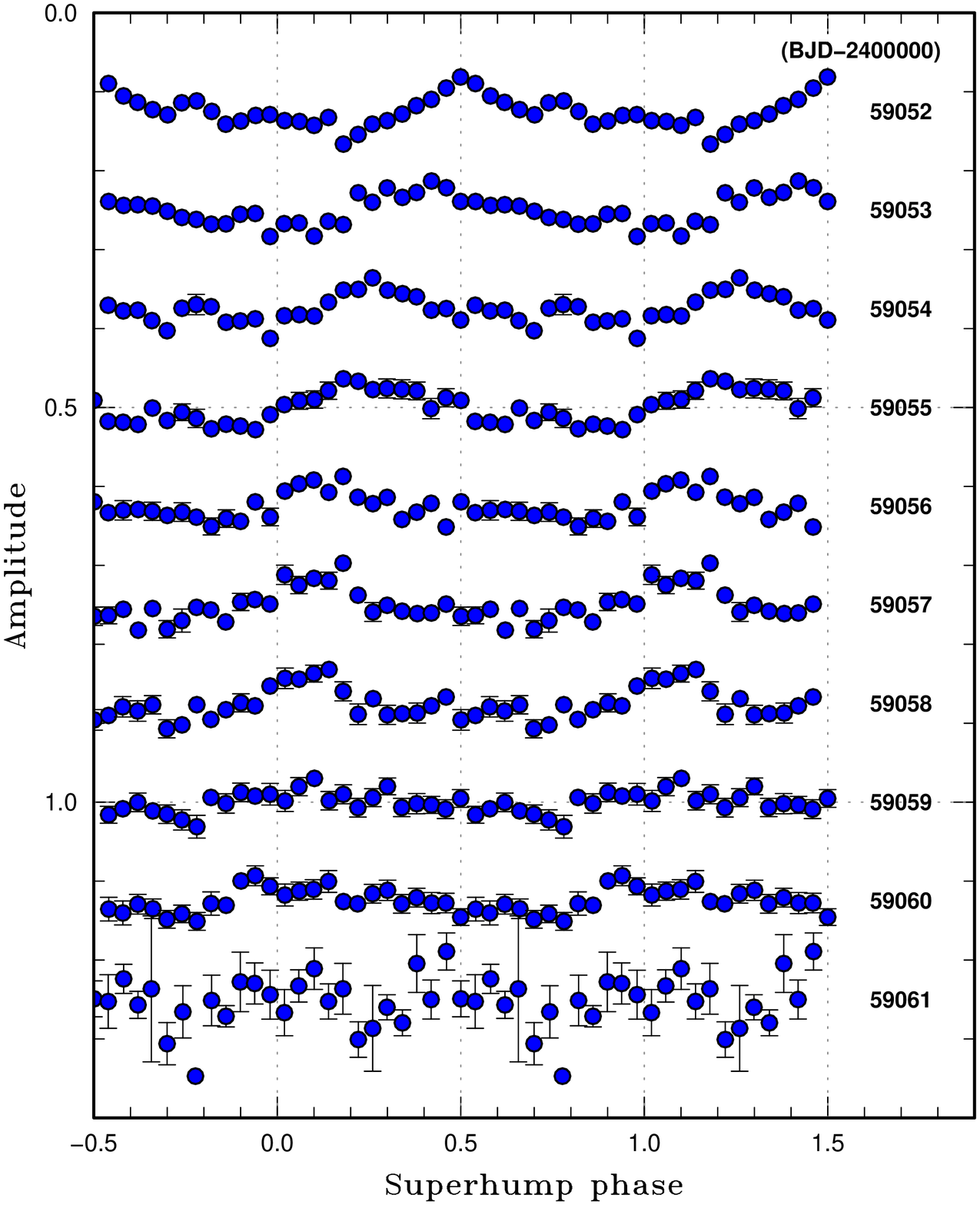}
\caption{
  Variation of profiles of superhumps in the late part of
  stage B.  1-d segments were used whose centers are shown on the
  right side of the figure.
  The zero phase and superhump period were
  the same as in figure \ref{fig:prof}.
  Superhumps decayed and the period became longer than in
  figure \ref{fig:prof2}.
}
\label{fig:prof3}
\end{center}
\end{figure*}

   The variation of the superhump profile is shown in
figures \ref{fig:prof}, \ref{fig:prof2} and \ref{fig:prof3}.
Figure \ref{fig:prof} contains stage A and the early phase
of stage B.  The amplitudes of the superhumps grew quickly
and the period became shorter (early stage B).
Figures \ref{fig:prof2} and \ref{fig:prof3} show
the evolution in stage B.  The superhump amplitudes decayed
with lengthening of the superhump period, as shown in
the $O-C$ analysis.  No phase reversal was observed
as in ER UMa \citep{kat96erumaSH,kat03erumaSH}.
This result suggests diversity in evolution of superhumps
among ER UMa stars.

   This superoutburst showed brightening in the later phase
(subsection \ref{sec:longterm}).  In ordinary SU UMa stars,
stage C superhumps and an increase in the superhump amplitude
are usually associated with such brightening
(see e.g., \cite{Pdot,kat22v844her}).  It was unclear whether
the same thing happened in CM Mic due to the lack of time-resolved
photometric data.

\begin{table*}
\caption{Times of superhump maxima in CM Mic}\label{tab:shmax}
\begin{center}
\begin{tabular}{cccc|cccc|cccc}
\hline
$E$ & $T$\commenta & error & amp\commentb &
$E$ & $T$ & error & amp &
$E$ & $T$ & error & amp \\
\hline
\xxinput{tab1a.inc}
\hline
\multicolumn{12}{l}{\commenta BJD$-$2459000.} \\
\multicolumn{12}{l}{\commentb Amplitude (mag).} \\
\end{tabular}
\end{center}
\end{table*}

\addtocounter{table}{-1}

\begin{table*}
\caption{Times of superhump maxima in CM Mic (continued)}
\begin{center}
\begin{tabular}{cccc|cccc|cccc}
\hline
$E$ & $T$\commenta & error & amp\commentb &
$E$ & $T$ & error & amp &
$E$ & $T$ & error & amp \\
\hline
\xxinput{tab1b.inc}
\hline
\multicolumn{12}{l}{\commenta BJD$-$2459000.} \\
\multicolumn{12}{l}{\commentb Amplitude (mag).} \\
\end{tabular}
\end{center}
\end{table*}

\section{Long-term light curves of other ER UMa stars with standstills}\label{sec:other}

\subsection{DDE 48}

   DDE 48 is a dwarf nova discovered by D. Denisenko in 2016
(vsnet-alert 20146\footnote{
   $<$http://ooruri.kusastro.kyoto-u.ac.jp/mailarchive/vsnet-alert/20146$>$.
}).  This object was identified as an ER UMa star
based on the detection of superhumps with a period of
0.067~d and frequent normal outbursts
(D. Denisenko, vsnet-alert 20291\footnote{
   $<$http://ooruri.kusastro.kyoto-u.ac.jp/mailarchive/vsnet-alert/20291$>$.
}).  A superhump was also reported in \citet{Pdot9}.
TESS observations in 2022 August also recorded a superoutburst
(N. Kojiguchi in prep.).
The object was also classified as a Z Cam star by the detection
of a standstill in ZTF data (T. Kato, vsnet-chat 9059\footnote{
   $<$http://ooruri.kusastro.kyoto-u.ac.jp/mailarchive/vsnet-chat/9059$>$.
}).

\begin{figure*}
\begin{center}
\includegraphics[width=16cm]{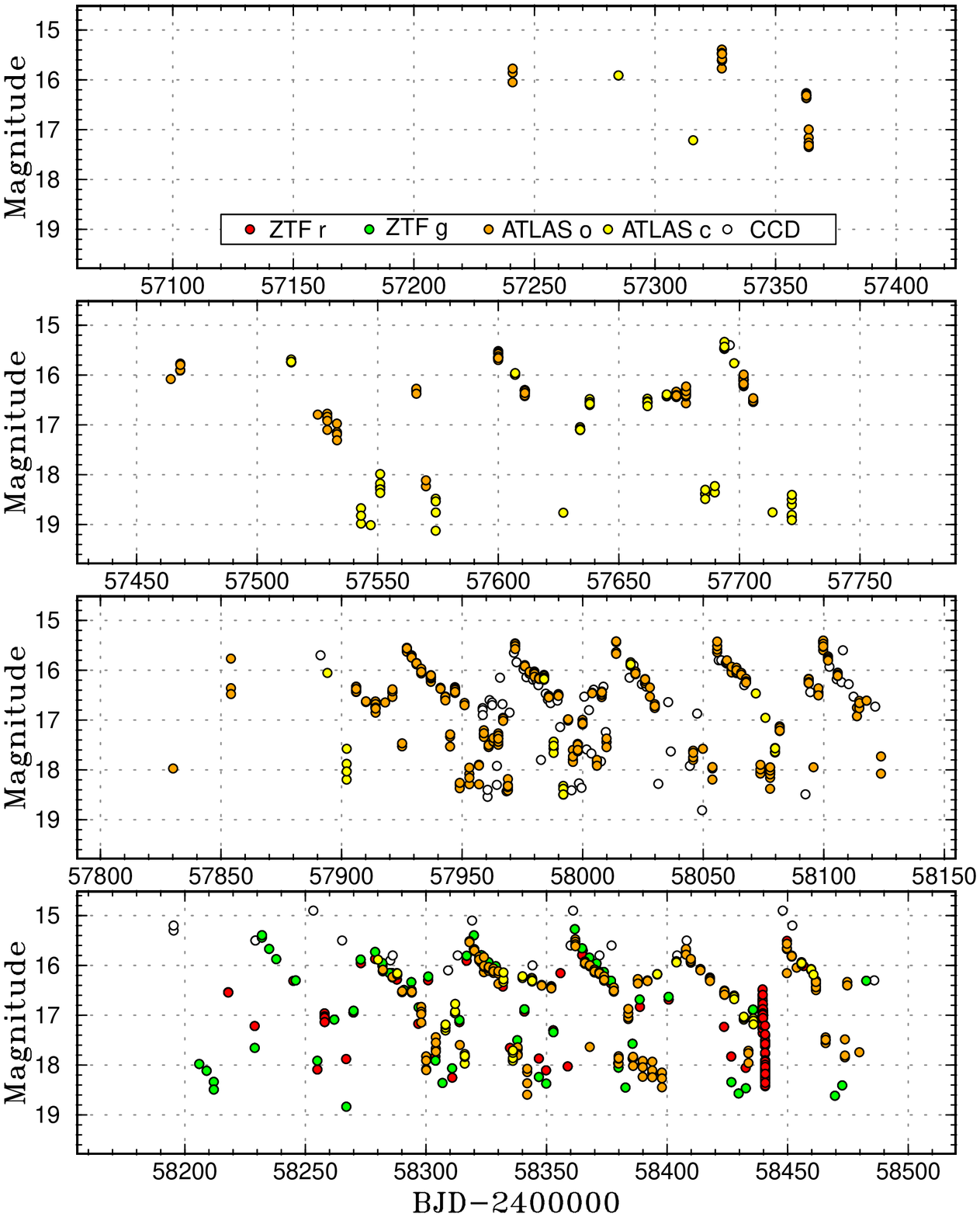}
\caption{
   Light curve of DDE 48 in 2015--2018.  The object mostly
showed ER UMa-type supercycles.
CCD refer to snapshot unfiltered CCD observations
reported to VSOLJ and VSNET.
}
\label{fig:dde48-lc1}
\end{center}
\end{figure*}

\begin{figure*}
\begin{center}
\includegraphics[width=16cm]{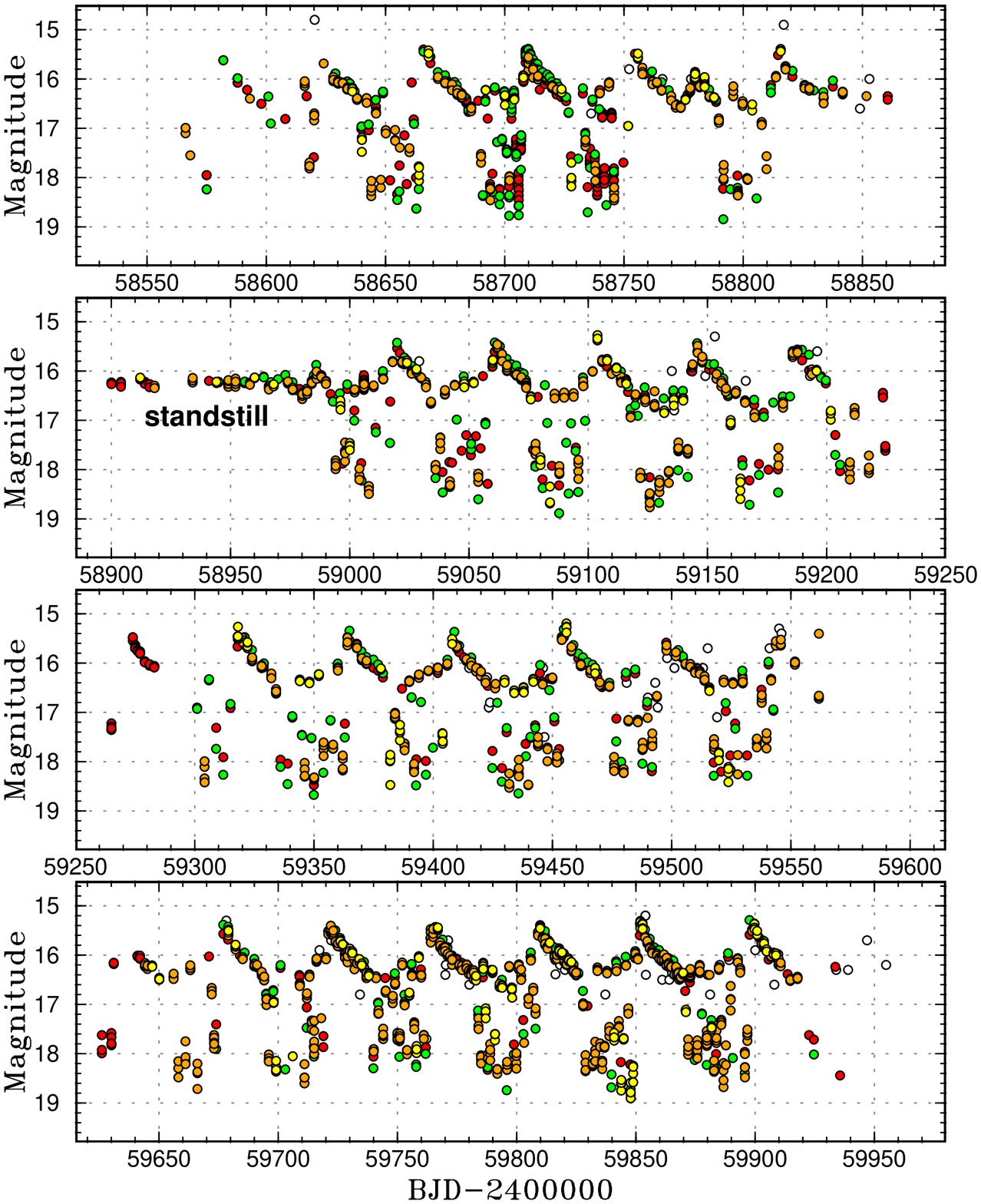}
\caption{
   Light curve of DDE 48 in 2019--2022.
The symbols are the same as in figure \ref{fig:dde48-lc1}.
There was a standstill between BJD 2458826 (2019 December 8)
and BJD 2458993 (2020 May 23) (end of the first to the start
of the second panels) in addition to ER UMa-type supercycles.
}
\label{fig:dde48-lc2}
\end{center}
\end{figure*}

   The long-term light curve is shown in figures \ref{fig:dde48-lc1}
and \ref{fig:dde48-lc2}.  The object showed ER UMa-type
supercycles most of the time.  There was a standstill between
BJD 2458826 (2019 December 8) and BJD 2458993 (2020 May 23).
The two superoutbursts preceding this standstill had
longer durations than the typical ones in this system
(first panel of figure \ref{fig:dde48-lc1}) and there
was a shallow dip amid the superoutburst plateau
(BJD 2458770--2458777) just like in some superoutbursts
in CM Mic.  The lengthening of superoutbursts toward
the standstill was probably a manifestation of the increasing
mass-transfer rate.
Just before the end of the standstill, small-amplitude
oscillations appeared.  This probably reflected a part of
the disk becoming thermally unstable due to
the decreasing mass-transfer rate.
This object is also known as
PS1-3PI J204611.81$+$242057.2 \citep{ses17PS1RR},
ATO J311.5492$+$24.3492 \citep{hei18ATLASvar} and
a Gaia variable Gaia DR3 1843254106354614272 (type CV)
\citep{GaiaDR3}.

\subsection{MGAB-V728}

\begin{figure*}
\begin{center}
\includegraphics[width=16cm]{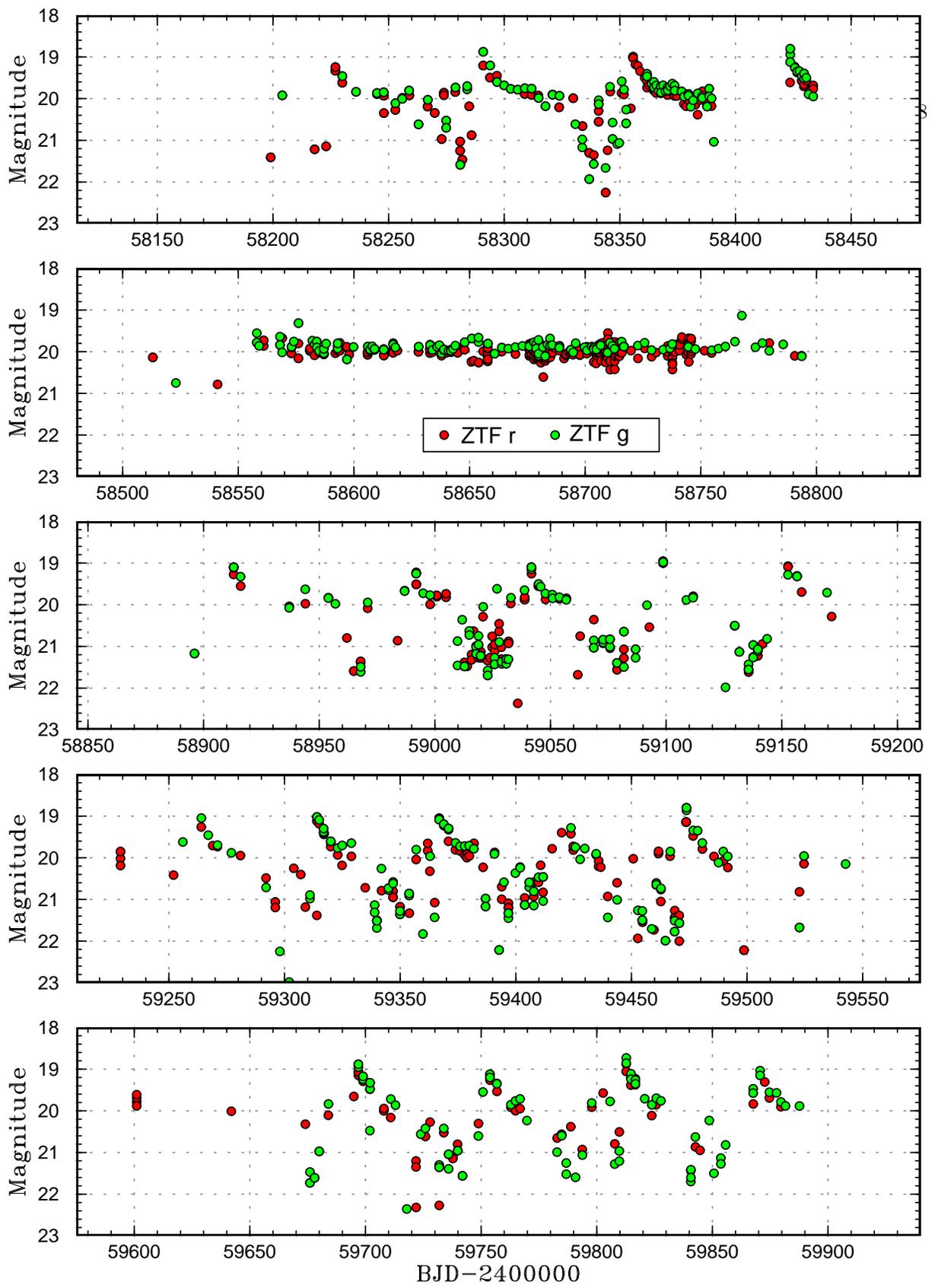}
\caption{
   Light curve of MGAB-V728 in 2018--2022.  The object
showed ER UMa-type supercycles except a standstill in 2019
(second panel).
}
\label{fig:mgabv728-lc1}
\end{center}
\end{figure*}

   MGAB-V728 was discovered as a Z Cam star by G. Murawski
in 2019.\footnote{
   $<$https://sites.google.com/view/mgab-astronomy/mgab-v701-v750$>$.
   This page has been updated after the initial submission to
the AAVSO VSX \citep{wat06VSX}.
}
The ZTF light curve is shown in figure \ref{fig:mgabv728-lc1}.
A standstill was present in 2019.  At other times, the object
displayed ER UMa-type behavior.  This object was also
listed as a Gaia variable Gaia DR3 4530506601048215296
(type CV) \citep{GaiaDR3}.  Superhumps have not yet been confirmed.

\subsection{MGAB-V3488}

\begin{figure*}
\begin{center}
\includegraphics[width=16cm]{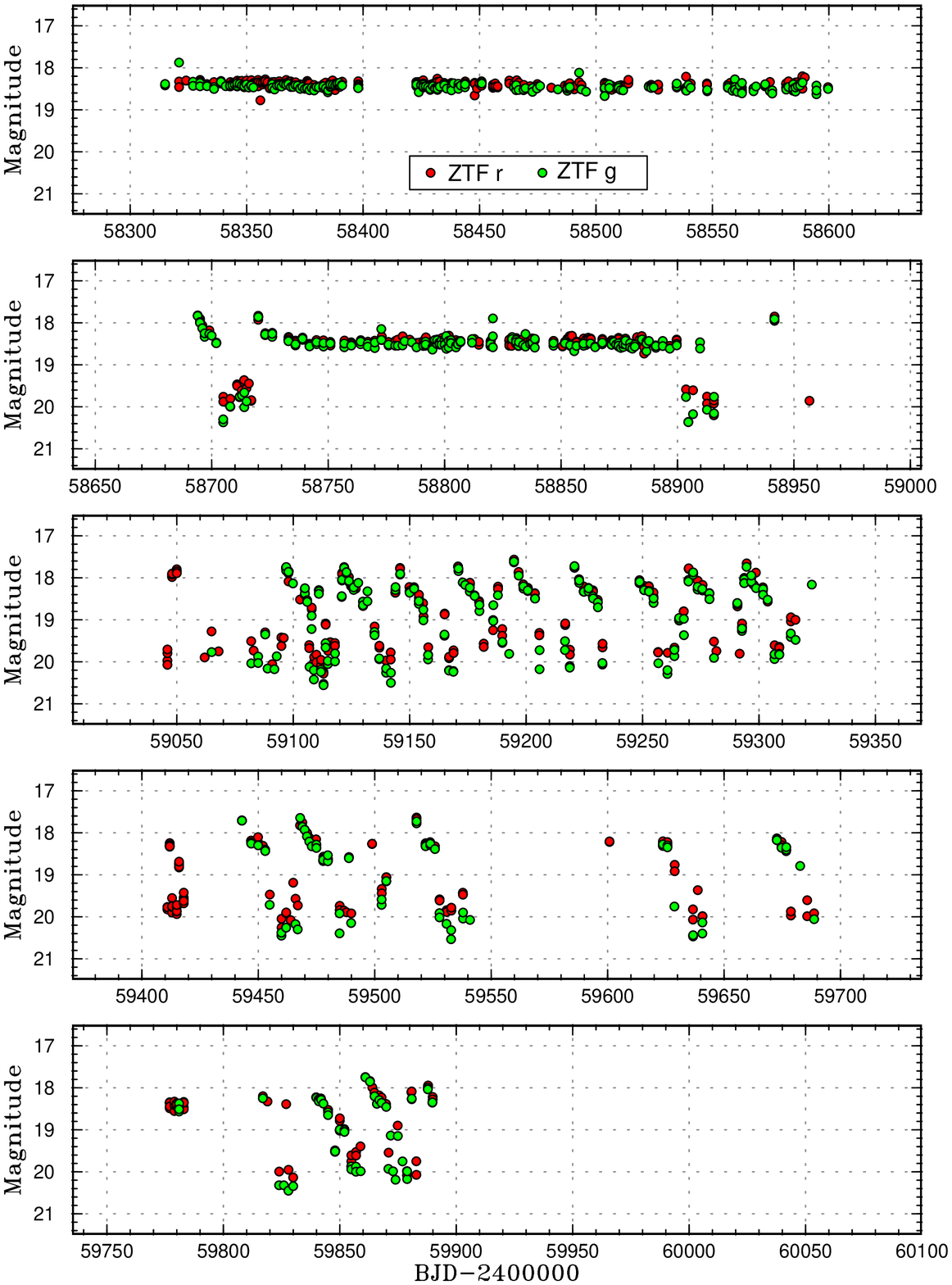}
\caption{
   Light curve of MGAB-V3488 in 2018--2022.  The object
was in standstill at least up to 2019 April (first panel).
The object showed an ER UMa-type supercycle between
BJD 2458694 (2019 July 29) and 2458720 (2019 August 24)
(second panel), then entered a standstill again.
In 2020--2022, the object was mostly in ER UMa states
with short ($\sim$25~d) supercycles. 
}
\label{fig:mgabv3488-lc1}
\end{center}
\end{figure*}

   This object was originally selected as a candidate RR Lyr
star (PS1-3PI J045544.81$+$653834.2: \cite{ses17PS1RR}).
G. Murawski (MGAB-V3488\footnote{
   $<$https://sites.google.com/view/mgab-astronomy/mgab-v3451-v3500$>$.
} reported it to be a Z Cam star.  M. Bajer also reported
its variability (BMAM-V809).
The object was identified as an ER UMa star with a short
supercycle ($\sim$25~d) comparable to RZ LMi (T. Kato,
vsnet-chat 8778\footnote{
   $<$http://ooruri.kusastro.kyoto-u.ac.jp/mailarchive/vsnet-chat/8778$>$.
}).  The ZTF light curve us shown in figure \ref{fig:mgabv3488-lc1}.
The object was in standstill at least up to 2019 April.
The object showed an ER UMa-type supercycle between
BJD 2458694 (2019 July 29) and 2458720 (2019 August 24);
normal outbursts were not recorded due to the sparse coverage,
then entered a standstill again.  This standstill ended
by fading (BJD 2458903 = 2020 February 24).
In 2020--2022, the object was mostly in ER UMa states
with short ($\sim$25~d) supercycles.
There was a possible standstill around BJD 2459777
(2022 July 16) to BJD 2459783 (2022 July 22), which was
not well covered by observations.
The ER UMa-type classification of the object was confirmed
by the presence of short outbursts between longer ones
(superoutbursts).
Superhumps have not yet been confirmed.

\subsection{PS1-3PI J181732.65$+$101954.6}

\begin{figure*}
\begin{center}
\includegraphics[width=16cm]{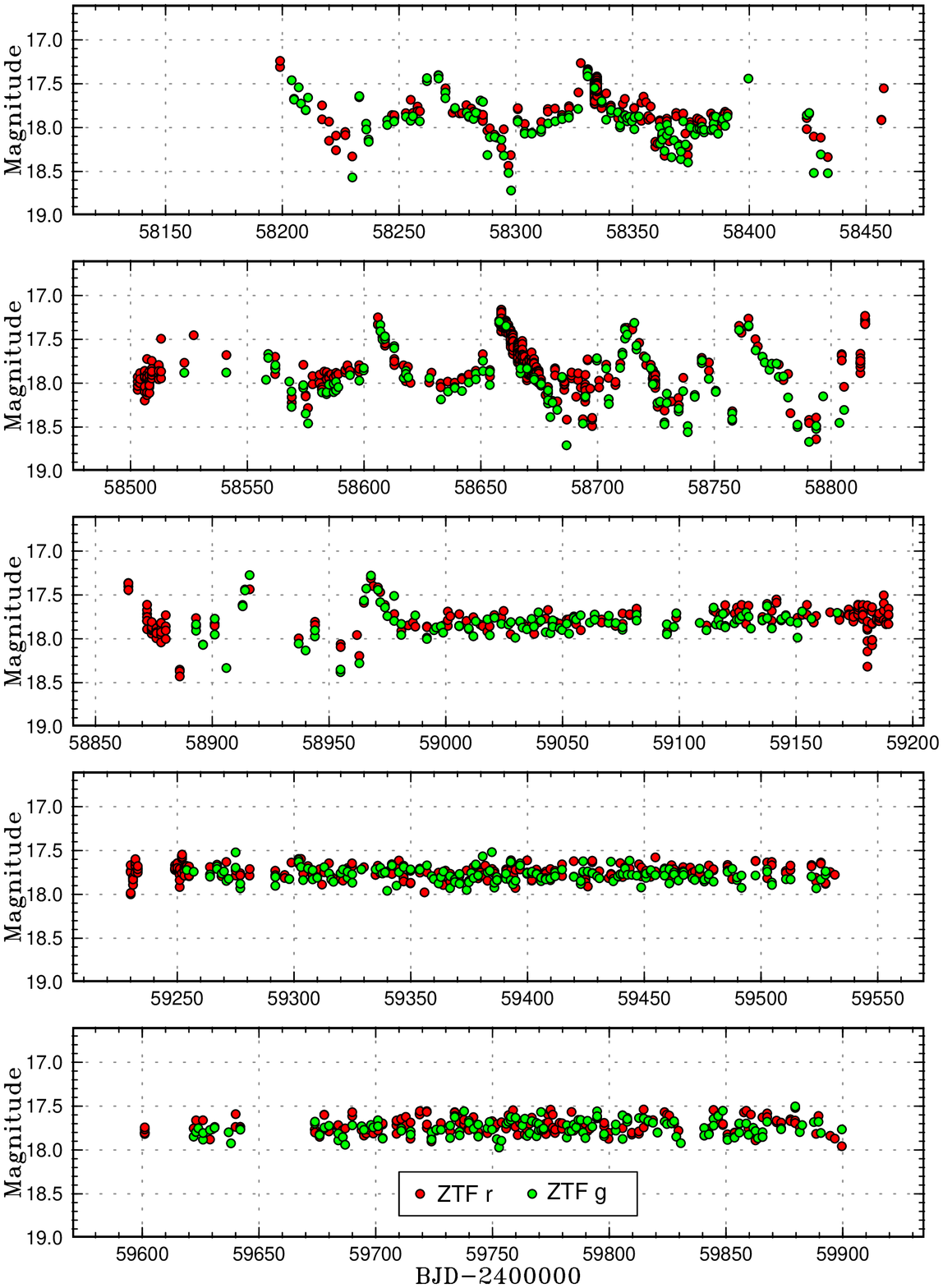}
\caption{
   Light curve of PS1-3PI J181732.65$+$101954.6 in 2018--2022.
The object was in an ER UMa state in 2019 to 2020 May
(second and third panels), then entered a long standstill
lasting up to now.  The behavior in 2018 (first panel)
was unusual and it looked like an IW And star (see text).
}
\label{fig:j181732-lc1}
\end{center}
\end{figure*}

\begin{figure*}
\begin{center}
\includegraphics[width=14cm]{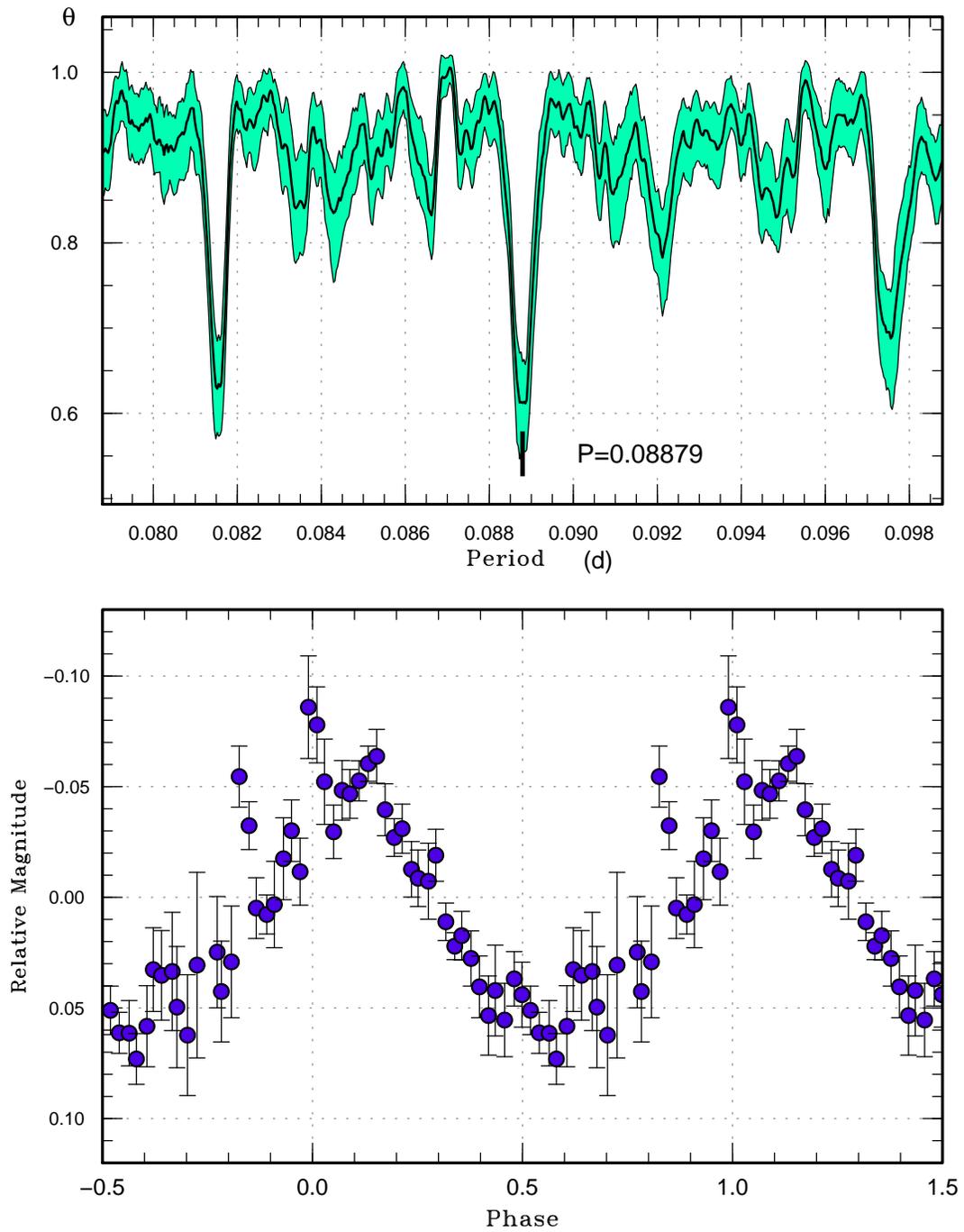}
\caption{
   Mean superhump profile of PS1-3PI J181732.65$+$101954.6
   from ZTF time-resolved data for BJD 2458657.8--2458674.8.
   (Upper): PDM analysis.  See figure \ref{fig:pdm} for explanation.
   (Lower): Phase plot.
}
\label{fig:j181732-pdm}
\end{center}
\end{figure*}

   This object was selected as a candidate RR Lyr
star \citep{ses17PS1RR}.  T. Kato found it to be a dwarf nova
(vsnet-chat 8487\footnote{
  $<$http://ooruri.kusastro.kyoto-u.ac.jp/mailarchive/vsnet-chat/8487$>$.
}).  It was originally considered as an SS Cyg star, but
later found to be an ER UMa star which entered a standstill
in 2020 May (T. Kato, vsnet-alert 26776\footnote{
  $<$http://ooruri.kusastro.kyoto-u.ac.jp/mailarchive/vsnet-alert/26776$>$.
}.  The ZTF light curve is shown in figure \ref{fig:j181732-lc1}.
As stated in vsnet-alert 26776, the behavior in 2018
(at least up to BJD 2458358 = 2018 August 27)
was atypical and looked like repetitive rising standstills
followed by dips, which are characteristics of an IW And star
\citep{kat19iwandtype}.  Although there were time-resolved
ZTF data during one of brightening in 2018 (BJD 2458333--2458334
= 2018 August 3--4), regular superhumps were not recorded.
We therefore consider that the behavior in 2018 was not
a variety of superoutbursts in an ER UMa star.
The behavior, however, did not perfectly
match that of a typical IW And star in that it did not show
brightening just before a dip.
There was a superoutburst in 2019 June--July, which
was not clearly preceded by a dwarf nova state and it looked
like brightening from a plateau phase of the preceding
outburst.  Superhumps were recorded during this 2019 June--July
outburst by ZTF time-resolved photometry
(figure \ref{fig:j181732-pdm}).  The mean superhump period
during this interval was 0.08879(2)~d.
This object is also known as BMAM-V769 (M. Bajer)\footnote{
  $<$https://www.aavso.org/vsx\_docs/2213875/3635/BMAM-V769\%20light\%20curve.PNG$>$.
} and ZTF J181732.64$+$101954.5 \citep{ofe20ZTFvar}.

\subsection{ZTF18abmpkbj}

\begin{figure*}
\begin{center}
\includegraphics[width=16cm]{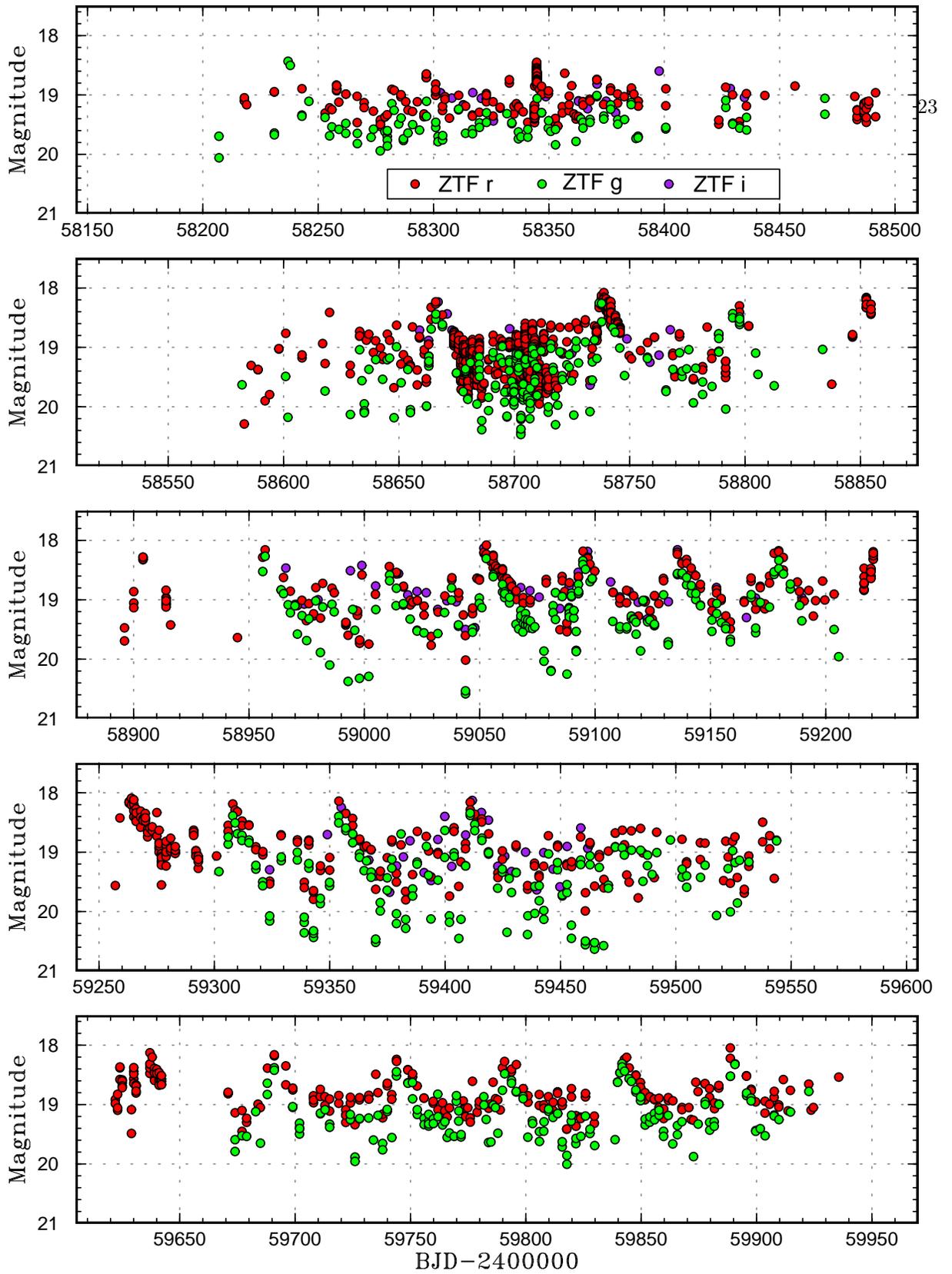}
\caption{
   Light curve of ZTF18abmpkbj in 2018--2022.
In addition to ER UMa states, two standstills
in 2018 (first panel) and 2021 September--November
(fourth panel) were present.
}
\label{fig:ztf18abmpkbj-lc1}
\end{center}
\end{figure*}

   This object was originally selected as a candidate RR Lyr
star (PS1-3PI J202831.22$+$200027.3) \citep{ses17PS1RR}.
The object was found to be an ER UMa star (T. Kato, vsnet-chat 8649\footnote{
  $<$http://ooruri.kusastro.kyoto-u.ac.jp/mailarchive/vsnet-chat/8649$>$.
}).  The classification was updated later
(T. Kato, vsnet-chat 9037\footnote{
  $<$http://ooruri.kusastro.kyoto-u.ac.jp/mailarchive/vsnet-chat/9037$>$.
}) based on a standstill starting in 2021 September.
This object was listed as a variable star ZTF J202831.22$+$200027.2
\citep{ofe20ZTFvar}.
The ZTF light curve is shown in figure \ref{fig:ztf18abmpkbj-lc1}.
In addition to ER UMa states, two standstills
in 2018 and 2021 September--November were present.
The ER UMa-type supercycle starting on BJD 2458664 (2019 June 29)
had many normal outbursts between the superoutbursts
and the supercycle was long (71~d).  The behavior became more
regular between BJD 2459052 (2020 July 21) and 2459411 (2021 July 15)
showing typical ER UMa-type supercycles with 42--56~d.
More variations were recorded during the 2021 standstill
than in the 2018 one and the former may have shown weak
dwarf nova-type outbursts during the standstill.
The object was again in a typical ER UMa state in 2022.
Superhumps have not yet been confirmed.

\subsection{ZTF18abncpgs}

\begin{figure*}
\begin{center}
\includegraphics[width=16cm]{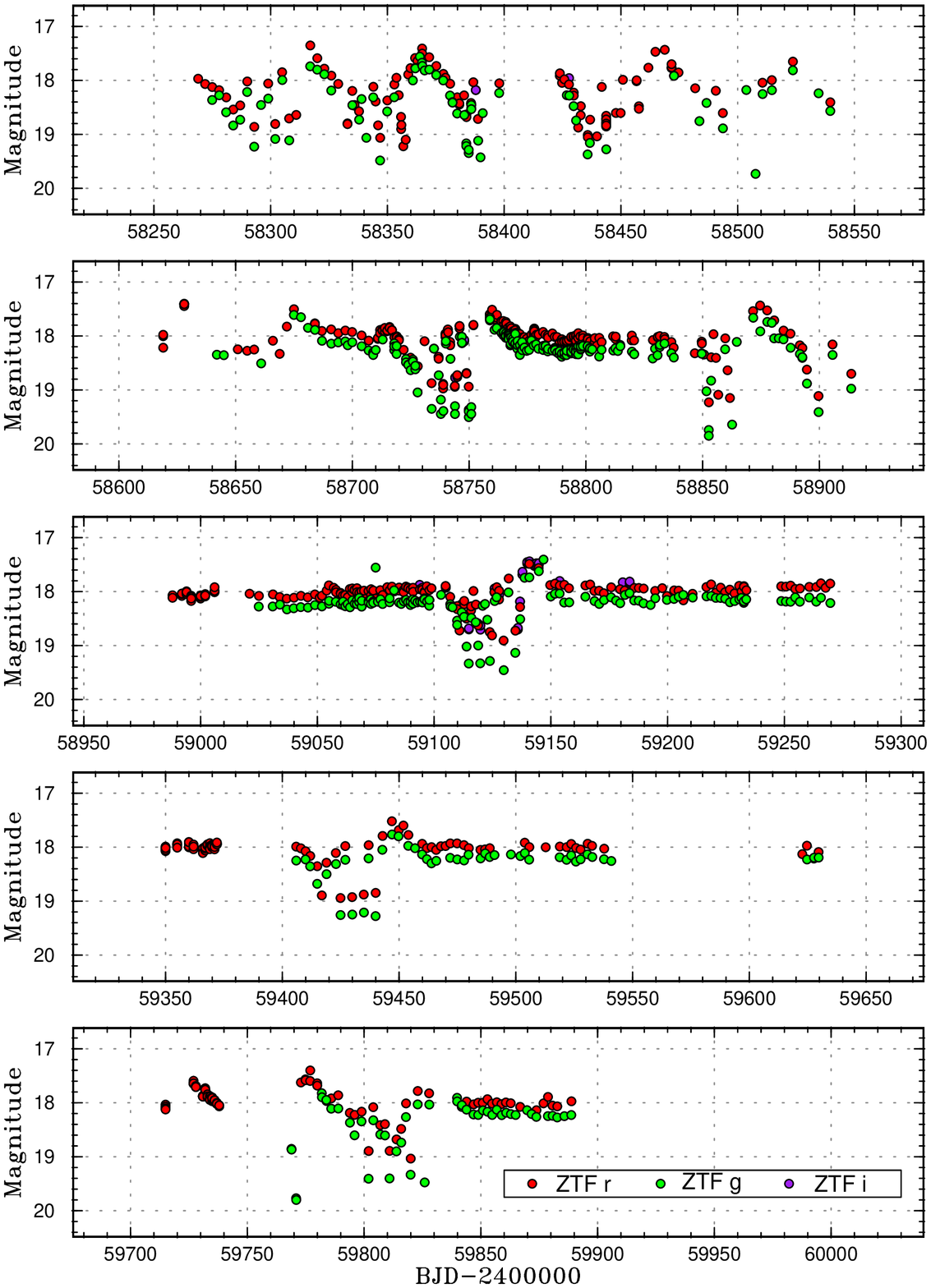}
\caption{
   Light curve of ZTF18abncpgs in 2018--2022.
Since 2019, the object stayed in standstill much of the time
while there were occasional ER UMa states.
}
\label{fig:ztf18abncpgs-lc1}
\end{center}
\end{figure*}

   This object was originally selected as a candidate RR Lyr
star (PS1-3PI J005932.18$+$570342.7) \citep{ses17PS1RR}.
AAVSO VSX \citep{wat06VSX} listed the object as a Z Cam star
in 2021.  T. Kato (vsnet-chat 8896\footnote{
  $<$http://ooruri.kusastro.kyoto-u.ac.jp/mailarchive/vsnet-chat/8896$>$.
}) noted that this object is similar to the ER UMa star RZ LMi
with variable supercycles in addition to a standstill.
The ZTF light curve is shown in figure \ref{fig:ztf18abncpgs-lc1}.
Since 2019, the object stayed in standstill much of the time
while there were occasional ER UMa states.
In 2018, this object was apparently in an ER UMa state
with a mean supercycle of $\sim$52~d.
The superoutburst starting on BJD 2458672 (2019 July 19)
showed brightening in the later phase.  The next superoutburst
had a long plateau (or a short standstill) as seen in
RZ LMi \citep{kat16rzlmi}.  The later standstill after
2020 May was interrupted by one or two ER UMa-type dwarf nova
cycles.
This object is also known as ATO J014.8840$+$57.0618 \citep{hei18ATLASvar}
and ZTF J005932.18$+$570342.7 \citep{ofe20ZTFvar}.
Superhumps have not yet been confirmed.

\subsection{ZTF19aarsljl}

\begin{figure*}
\begin{center}
\includegraphics[width=16cm]{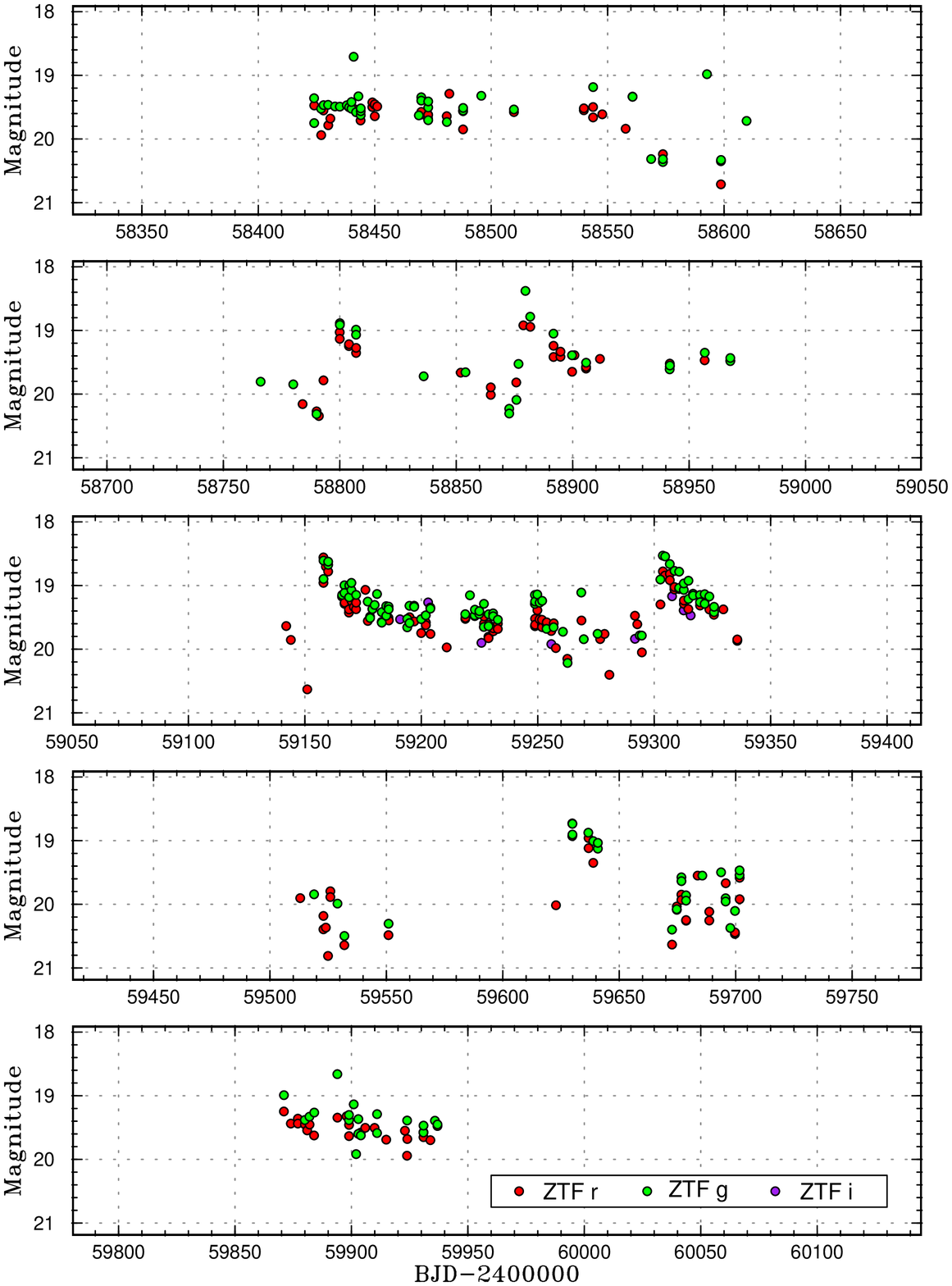}
\caption{
   Light curve of ZTF19aarsljl in 2018--2022.
The part BJD 2459155 (2020 November 4) to BJD 2459304
(2021 March 30, third panel) looks like an extended superoutburst
followed by dwarf nova outbursts as in RZ LMi \citep{kat16rzlmi}.
The short cycle between BJD 2458800 (2019 November 12) and
2458880 (2020 January 31, second panel) suggests an ER UMa-type object.
The segment between BJD 2458424 (2018 November 1) and
2458560 (2019 March 18, first panel) was apparently
a standstill.
}
\label{fig:ztf19aarsljl-lc1}
\end{center}
\end{figure*}

   This object was originally selected as a candidate RR Lyr
star (PS1-3PI J084907.10$-$152526.7) \citep{ses17PS1RR}.
T. Kato (vsnet-chat 9068\footnote{
  $<$http://ooruri.kusastro.kyoto-u.ac.jp/mailarchive/vsnet-chat/9068$>$.
}) noted that this object is likely an ER UMa star with
standstills.
The ZTF light curve is shown in figure \ref{fig:ztf19aarsljl-lc1}.
The part BJD 2459155 (2020 November 4) to BJD 2459304
(2021 March 30) looks like an extended superoutburst
followed by dwarf nova outbursts as in RZ LMi \citep{kat16rzlmi}.
The short cycle between BJD 2458800 (2019 November 12) and
2458880 (2020 January 31) suggests an ER UMa-type object.
The segment between BJD 2458424 (2018 November 1) and
2458560 (2019 March 18) was apparently
a standstill.  Although the quality of the data was not
sufficient due to the faintness of the object and the short
coverage by ZTF, we consider that this object is an ER UMa
star close to the thermal stability as in RZ LMi when it
showed standstills.
This object was also listed as a Gaia variable
Gaia DR3 5733518178924667392 (type CV) \citep{GaiaDR3}.
Superhumps have not yet been confirmed.

\subsection{MGAB-V284}

\begin{figure*}
\begin{center}
\includegraphics[width=16cm]{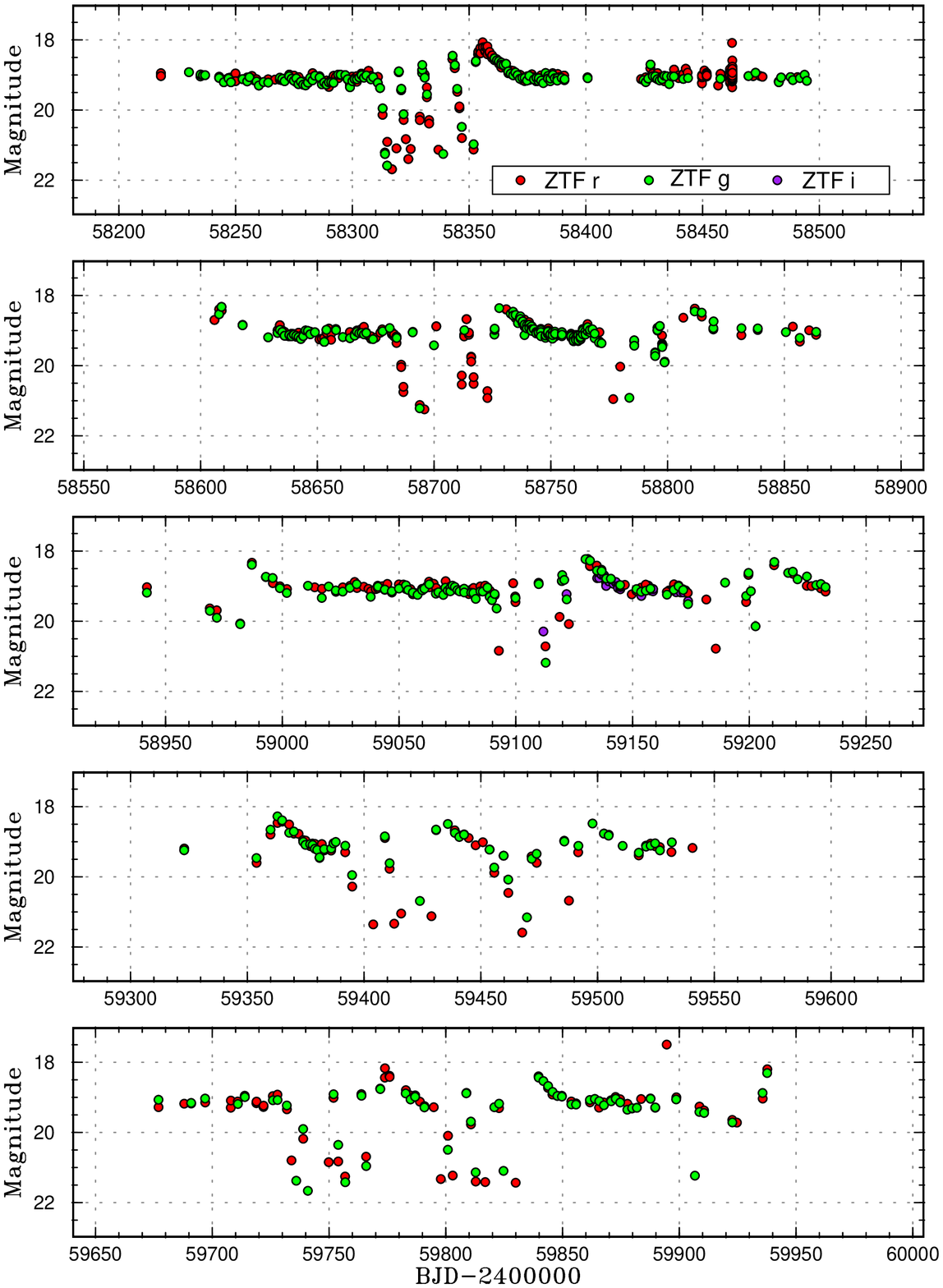}
\caption{
   Light curve of MGAB-V284 in 2018--2022.
}
\label{fig:mgabv284-lc1}
\end{center}
\end{figure*}

\begin{figure*}
\begin{center}
\includegraphics[width=16cm]{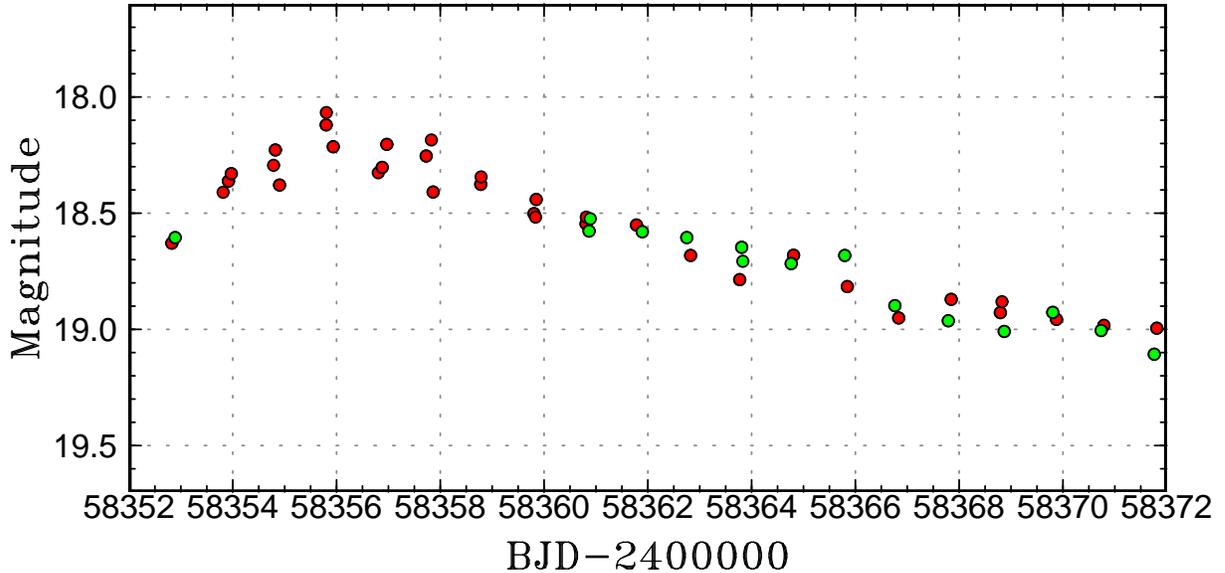}
\caption{
   Light curve of MGAB-V284 around the peak of the 2018 October
outburst.
The symbols are the same as in figure \ref{fig:mgabv284-lc1}.
}
\label{fig:mgabv284-lclarge}
\end{center}
\end{figure*}

   This object was reported as a Z Cam star by G. Murawski.\footnote{
   $<$https://www.aavso.org/vsx\_docs/844556/3120/MGAB-V284.png$>$.
}
T. Kato pointed out in vsnet-alert 23716\footnote{
  $<$http://ooruri.kusastro.kyoto-u.ac.jp/mailarchive/vsnet-alert/23716$>$.
} as follows (typos corrected):
``According to ZTF data, a large and long outburst
triggered a standstill.  The outburst amplitudes between
the standstills gradually grew up.
While the outburst amplitude just after the end of the standstill
is 2.5~mag and the duration is 3~d, the outburst amplitude just
before the standstill is 3.5~mag and the duration is 20~d.
In addition, recurrence time might be 120~d
according to Apparent Magnitude on Lasair data
(https://lasair.roe.ac.uk/object/ZTF18abgpmcd/)''.
This statement was based on ZTF Lasair data \citep{ZTFLasair}
and is updated here using the most recent release of the ZTF data
(figure \ref{fig:mgabv284-lc1}).
All ``outbursts'' occurred after a sequence of short-period
dwarf nova oscillations and the behavior is most analogous
to ER UMa stars with long standstills, such as ZTF18abncpgs.
The only difference is in the time scale: the cycle lengths
of normal outbursts are $\sim$10~d compared to $\sim$4~d in
ER UMa stars.  This difference implies that the orbital period
is longer than those of typical ER UMa stars. 
In recent years, however, a number of SU UMa stars above
the period gap have been identified, breaking the general
consensus \citep{war95book} such as BO Cet \citep{kat21bocet,kat23bocet},
MisV1448 (N. Kojiguchi et al. in preparation),
ASASSN-18aan \citep{wak21asassn18aan} and
ASASSN-15cm \citep{kat23asassn15cm}, and MGAB-V284 may be
an ER UMa star with standstills above the period gap.
Although the period was not determined due to the very small
number of observations, the large scatter in the ZTF light curve
around the peak of the 2018 October outburst was suggestive
of superhumps (figure \ref{fig:mgabv284-lclarge}).
Further observations around the peak of a future bright
outburst are requested.
This object is also known as ZTF J221257.97$+$490644.4 \citep{ofe20ZTFvar}
and was also listed as a Gaia variable Gaia DR3 1975922107078810112
(type CV) \citep{GaiaDR3}.

\section{Summary}

   We analyzed ASAS-SN, ATLAS and TESS observations of CM Mic
and found that this object belongs to a small group of
ER UMa stars showing standstills.
The object showed typical ER UMa-type cycles between
2019 July and 2021 November, and in 2023.  It showed
standstills between 2017 and 2019 July, and in 2022.
The supercycles varied between 49 and 83~d, which probably
reflected the variable mass-transfer rate.
In 2015, the object showed outbursts with a cycle length
of $\sim$35~d in 2015 whose variations became weaker in 2016.
An analysis of TESS observations during the 2020 July outburst
detected superhumps with a mean period of 0.080251(6)~d
(value after the full development of superhumps).
and a period of the growing stage of 0.0817(2)~d.
The period derivative $+$2.0(2) $\times$ 10$^{-5}$ was
similar to those recorded in other ER UMa stars.

   We also studied previously undocumented ER UMa stars
showing standstills mainly using ZTF data.
\begin{itemize}
\item
DDE 48 mostly showed ER UMa-type supercycles but showed
a standstill in 2019 December--2020 May.
MGAB-V728 also mostly showed ER UMa-type supercycles but
showed a standstill in 2019.
\item
MGAB-V3488 showed long standstills in 2018--2020, but
was mostly in ER UMa states with short ($\sim$25~d)
supercycles in 2020--2022 similar to RZ LMi.
\item
PS1-3PI J181732.65$+$101954.6 showed ER UMa-type supercycles
up to 2020 May and entered a long standstill lasting
up to now.
\item
ZTF18abmpkbj mostly showed ER UMa-type supercycles
but two standstills were recorded in 2018 and late 2021.
These standstills had larger variations than in other
objects and this object may have shown weak dwarf nova-type
outbursts during the standstills.
\item
ZTF18abncpgs showed standstills most of the time,
but also showed ER UMa-type supercycles in 2018 and
occasionally between standstills.
\item
ZTF19aarsljl is likely an ER UMa star with standstills,
but the details were not clear due to the faintness.
\item
MGAB-V284 showed a pattern similar to ER UMa stars
showing standstills but with a longer time-scale of
normal outbursts.  We suspect that this object is
an ER UMa star with standstills above the period gap.
\end{itemize}
None of the objects we studied showed a superoutburst
arising from a long standstill, as recorded in NY Ser in 2018,
although the 2019 June--July superoutburst of
PS1-3PI J181732.65$+$101954.6 may have directly occurred
without experiencing a dwarf nova state and it could have
been analogous to a superoutburst arising from a standstill.
The case of NY Ser appears to be very rare and
we could not find additional evidence that the disk radius
during standstills in ER UMa stars increases.

\section*{Acknowledgements}

This work was supported by JSPS KAKENHI Grant Number 21K03616.
The authors are grateful to the ASAS-SN, ATLAS, TESS and ZTF teams
for making their data available to the public.
We are grateful to VSOLJ and VSNET observers (particularly
M. Moriyama, G. Poyner, Y. Maeda and M. Hiraga) who reported
snapshot CCD photometry of DDE 48.

Funding for the TESS mission is provided by the NASA Science
Mission Directorate.

This work has made use of data from the Asteroid Terrestrial-impact
Last Alert System (ATLAS) project.
The ATLAS project is primarily funded to search for
near earth asteroids through NASA grants NN12AR55G, 80NSSC18K0284,
and 80NSSC18K1575; byproducts of the NEO search include images and
catalogs from the survey area. This work was partially funded by
Kepler/K2 grant J1944/80NSSC19K0112 and HST GO-15889, and STFC
grants ST/T000198/1 and ST/S006109/1. The ATLAS science products
have been made possible through the contributions of the University
of Hawaii Institute for Astronomy, the Queen's University Belfast, 
the Space Telescope Science Institute, the South African Astronomical
Observatory, and The Millennium Institute of Astrophysics (MAS), Chile.

Based on observations obtained with the Samuel Oschin 48-inch
Telescope at the Palomar Observatory as part of
the Zwicky Transient Facility project. ZTF is supported by
the National Science Foundation under Grant No. AST-1440341
and a collaboration including Caltech, IPAC, 
the Weizmann Institute for Science, the Oskar Klein Center
at Stockholm University, the University of Maryland,
the University of Washington, Deutsches Elektronen-Synchrotron
and Humboldt University, Los Alamos National Laboratories, 
the TANGO Consortium of Taiwan, the University of 
Wisconsin at Milwaukee, and Lawrence Berkeley National Laboratories.
Operations are conducted by COO, IPAC, and UW.

The ztfquery code was funded by the European Research Council
(ERC) under the European Union's Horizon 2020 research and 
innovation programme (grant agreement n$^{\circ}$759194
-- USNAC, PI: Rigault).

\section*{List of objects in this paper}
\xxinput{objlist.inc}

\section*{References}

We provide two forms of the references section (for ADS
and as published) so that the references can be easily
incorporated into ADS.

\newcommand{\noop}[1]{}\newcommand{\hyphalt}{-}

\renewcommand\refname{\textbf{References (for ADS)}}

\xxinput{cmmicaph.bbl}

\renewcommand\refname{\textbf{References (as published)}}

\xxinput{cmmic.bbl.vsolj}

\end{document}